\documentclass[floatfix,%
 reprint,
nofootinbib,
 amsmath,amssymb,
 aps,
floatfix,
]{revtex4-2}
\usepackage{subcaption}
\usepackage{caption}
\usepackage{graphicx}
\usepackage{siunitx}
\usepackage{braket}
\usepackage{subcaption}
\usepackage{makecell}
\usepackage{xcolor}
\usepackage{framed}
\usepackage{dcolumn}
\usepackage{bm}
\usepackage{float}
\usepackage{microtype}
\usepackage{mathrsfs}
\usepackage[colorlinks,citecolor=blue,urlcolor=blue,linkcolor=blue]{hyperref}
\usepackage{booktabs}

\begin{document}
\preprint{APS/123-QED}
\title{Testing the Spacetime Geometry of Sgr A* with the Relativistic Orbit of S2 star}

\author
{Parth Bambhaniya$^{1}$}
\email{grcollapse@gmail.com}
\author{Preet Dalal$^{2}$}
\email{pdalal2003@gmail.com}
\author{Giovani H. Vicentin$^{1}$}
\email{giovani.vicentin@usp.br}
\author{Riccardo Della Monica$^{3}$}
\email{rdellamonica@tecnico.ulisboa.pt}
\author{Elisabete M. de Gouveia Dal Pino$^{1}$}
\email{dalpino@iag.usp.br}
\author{Bina Patel$^{2}$}
\email{binapatel.maths@charusat.ac.in}

\affiliation{$^{1}$Instituto de Astronomia, Geofísica e Ciências Atmosféricas, Universidade de São Paulo, IAG, Rua do Matão 1225, CEP: 05508-090 São Paulo - SP - Brazil}

\affiliation{$^{2}$P. D. Patel Institute for Applied Sciences, Charusat University, Anand-388421, Gujarat, India,
}

\affiliation{$^{3}$CENTRA, Departamento de Física, Instituto Superior Técnico – IST, Universidade de Lisboa – UL, Avenida Rovisco Pais 1, 1049-001 Lisboa, Portugal}

\date{\today}

\begin{abstract}
In this work, we perform a relativistic test of the spacetime geometry of Sagittarius A* (Sgr A*) using the orbit of the S2 star. We consider a broad class
of compact object models, including Schwarzschild, Reissner-Nordström, Bardeen, Hayward, and Simpson-Visser black holes, as well as the Janis-Newman-Winicour naked singularity spacetime. For each geometry, we integrate the timelike geodesic equations and consistently project the resulting trajectories onto astrometric and spectroscopic observables, incorporating R{\o}mer time delay and relativistic redshift effects. The theoretical predictions are tested with current Very Large Telescope (VLT) observations of the S2 star, while
simultaneously imposing constraints from the Event Horizon Telescope shadow size. We find that several spacetimes that are degenerate at the level of shadow imaging, most notably Schwarzschild, Reissner-Nordström, and
Bardeen regular black hole geometries, remain statistically indistinguishable when tested against present S2 data. We further carry out a statistical model
comparison based on the Akaike and Bayesian information criteria (AIC and BIC) to evaluate the relative performance of the alternative spacetime models. Our analysis also constrains the generalized charge like parameter $q/M$ in
non-Schwarzschild spacetimes based on current S2 star observations, and
identifies specific black hole and horizonless geometries that can be further tested with forthcoming high precision astrometric observations from the VLT and Keck telescopes.

\vspace{.2cm}
$\boldsymbol{Key words}$ : S2/S0-2 star, Black Holes, Compact Objects, Precession, Orbits, Galactic Center.
\end{abstract}
\maketitle

\section{Introduction}

The center of our Galaxy presents a unique environment where gravity, astrophysics, and the causal structure of spacetime can be studied directly~\cite{deLaurentis:2022oqa}. At its heart lies Sgr A*, a compact object with a mass of approximately $4\times10^{6} M_{\odot}$, located at a distance of roughly $8$~kpc from Earth~\cite{Abuter2020}. Active, ongoing observational efforts by the Event Horizon Telescope (EHT)~\cite{EHTC2022}, the GRAVITY~\cite{Abuter2020}, and the UCLA Galactic Center group~\cite{Ghez2005,Hees2017} have turned the Galactic Center (GC) into the most precisely monitored region across the electromagnetic spectrum of strong gravitational fields. The relative proximity of Sgr A* allows for detailed measurements of the motions of stars orbiting close to the compact object, the structure and variability of the surrounding accretion flow, and multi-wavelength emission from the innermost regions~\cite{deLaurentis:2022oqa}. These high-precision observations have firmly established Sgr A* as the most reliably confirmed supermassive compact object in our Galaxy, providing an exceptional opportunity to probe the predictions of general relativity and to test alternative models for the nature of the central mass.

The S-stars, and in particular the S0-2/S2 star (hereafter S2), plays a central role in these tests. S2 has a short orbital period of about $\sim16$ years and a high eccentricity of nearly $0.88$~\cite{Abuter2020, Do2019}. Its pericenter passage brings it deep into the relativistic potential of Sgr A*. The orbit follows a timelike geodesic to very high accuracy. Its orbital precession, Doppler redshift, and orbital period probe the metric in the regime where post-Newtonian relativistic corrections become significant. The 2018 pericenter passage, during which S2 reached nearly $\sim 2.7\%$ of the speed of light at a distance of about $120$ AU (corresponding to $\sim1400$ gravitational radii of the central object), allowed the first direct measurement of gravitational redshift in this environment~\cite{GRAVITY:2018ofz, Do2019} and a precise determination of Schwarzschild precession in its subsequent motion~\cite{Abuter2020}. 

Several studies have utilized the orbits of S2 star to test alternative models of gravity theories. For instance, the geodesic motion of S2 star has been used to explore Scalar-Tensor-Vector Gravity~\cite{DellaMonica:2021xcf}, $f(R)$ gravity~\cite{deMartino:2021daj}, and scenarios involving wormhole~\cite{DellaMonica:2021fdr}. Other investigations have employed S2 star orbital data to constrain specific gravitational models~\cite{Tomaselli:2025zdo}. The parameters of $R^n$ gravity were limited in~\cite{Borka}, while Yukawa gravity parameters were constrained in~\cite{Borka:2013dba}. Graviton mass bounds comparable to those from LIGO-Virgo were obtained in~\cite{Zakharov:2016lzv,Zakharov:2018cbj}, and additional limits on tidal charge were derived in~\cite{Zakharov:2018awx}. More recently, Yukawa gravity constraints have been refined~\cite{Jovanovic:2022twh,Hees:2017aal,GRAVITY:2025ahf,Tan:2024wuk}, and graviton mass limits have been further improved using GRAVITY observations of S2~\cite{Jovanovic}. These studies highlight the growing precision of stellar dynamics as a tool to probe both general relativity and its alternatives in the GC. 

An alternative approach to modeling the central mass distribution has been proposed by Ruffini, Arg\"uelles, and Rueda (RAR), who introduced a self gravitating dark matter core halo configuration~\cite{Ruffini2015,Arguelles2022}. This model has been applied in several studies that claim that it can reproduce the orbits of bright stars around Sgr A* more accurately than the traditional supermassive black hole paradigm. In the RAR model, the dense dark matter core exhibits nearly constant density, producing elliptical stellar trajectories reminiscent of Keplerian motion. However, subsequent analysis by Zakharov~\cite{Zakharov:2021cgx} indicates that the RAR configuration predicts orbital periods that are largely independent of the semi-major axis, and that the centers of the elliptical orbits do not coincide with the GC. These features are in tension with current observational data, suggesting that while the RAR model provides a useful framework, it may not fully capture the dynamics of stars in the immediate vicinity of Sgr A*. Also current results show consistency with general relativity and place strong limits on any extended mass within the orbit of S2, at the level of only a few thousand solar masses~\cite{AbdElDayem2024}. On one hand, these constraints  rule out substantial concentrations of dark or stellar matter in the immediate vicinity of the compact object, on the other hand, they can be used to constrain dark matter models in the GC, one remarkable example being an ultralight dark matter soliton~\cite{DellaMonica:2022kow, DellaMonica:2023dcw}.

Although the Schwarzschild and Kerr metrics successfully describe many observed phenomena, it is not yet established that the compact object at the GC must be described by either solution. The Event Horizon Telescope has revealed a bright ring of emission on the scale of the photon orbit around Sgr~A*~\cite{EventHorizonTelescope:2022xqj}. While this observation is consistent with a Schwarzschild black hole, current uncertainties allow many distinct spacetimes to reproduce the measured shadow size. Detailed analyses show that several alternative metrics remain compatible with the EHT constraints ~\cite{Vagnozzi:2022moj}. These include Reissner-Nordstr\"om (RN)~\cite{Reissner1916} black hole, regular black holes such as the Simpson-Visser (SV)~\cite{Simpson:2018tsi}, Bardeen~\cite{Bardeen:2018frm}, and Hayward~\cite{Hayward:2005gi} geometries, as well as horizonless solutions like the Janis-Newman-Winicour (JNW)~\cite{Janis1968,Virbhadra:1997} and Joshi-Malafarina-Narayan (JMN-1) spacetimes~\cite{Vagnozzi:2022moj,Joshi2011,Saurabh:2023otl}. Shadow observables depend primarily on the null geodesic structure and therefore cannot uniquely determine the underlying causal geometry when different metrics share similar photon orbit properties~\cite{Bambhaniya:2025iqb}. Stellar dynamics provide an independent and essential probe of the GC spacetime, as they trace timelike geodesics and are sensitive to the gravitational field over a wide radial range, including relativistic periastron precession, gravitational redshift and time dilation effects, orbital stability, and the influence of effective matter sources in non-vacuum or regular spacetimes~\cite{Katsumata2024,Bambhaniya2019a,Harada2023,Bambhaniya2021a,Bambhaniya2022,Igata2023,Bambhaniya:2025xmu,Bambhaniya2021b,Crespi:2025ygl}.
These considerations motivate a unified investigation of both null and timelike geodesics in the GC spacetime. Any viable model of Sgr A* must simultaneously reproduce the observed shadow diameter and the detailed orbital properties of the S2 star. In this work, we present a systematic analysis of the S2 orbit within a representative set of compact object spacetimes. These include the RN black hole, the Bardeen, Hayward, and SV geometries, and the JNW naked singularity models. By confronting each metric with current astrometric and spectroscopic observations and imposing independent constraints from the EHT shadow size, we identify which spacetimes remain compatible with the data and which are disfavored. Stellar dynamics and horizon-scale imaging together provide a stringent test of the causal structure of the compact object at the center of our Galaxy.

We also note that the constraining power of the S2 orbit becomes even more limited when black hole spin is considered. A consistent treatment of frame-dragging effects requires a fully relativistic projection of the stellar trajectory from the orbital plane to the observer’s sky. As of our current knowledge, no such projection has been derived in a form suitable for direct fitting to current astrometric and spectroscopic data (except for a proof-of-concept sensitivity analysis presented in~\cite{Grould:2017bsw}). As a result, most existing analyses rely on Keplerian or post-Newtonian sky projections. This limitation significantly weakens the ability of present S2 data to place robust constraints on the Kerr spin parameter. For this reason, several observational analyses, including those by the UCLA, and GRAVITY collaborations, adopt non-spinning spacetime models when fitting S-star orbits.

This paper is organized as follows. In Sec.~\ref{sec:2}, we introduce the spacetime models considered in this work and describe their key geometric properties, including Schwarzschild, charged, regular, and horizonless compact object geometries. In Sec.~\ref{sec:3}, we discuss the associated energy conditions and their physical interpretation. The formulation of fully-relativistic stellar orbits and the adopted initial conditions are presented in Sec.~\ref{sec:4}. In Sec.~\ref{sec:5}, we describe the observational data, the construction of astrometric and spectroscopic observables, and the Bayesian inference framework used to estimate orbital parameters and compare models. In Sec.~\ref{sec:constraint}, we obtain the model parameter constraints using observational precession of S2 star. The results of our analysis are presented and discussed in Sec.~\ref{sec:results}, followed by the conclusions in Sec.~\ref{sec:conc}. Throughout this work, we use the metric signature $(-,+,+,+)$ and adopt astrophysical units with explicit factors of $G$ and $c$. 

\section{Spacetime Models}
\label{sec:2}
In this work, we explore several spacetime geometries that capture a wide range of relativistic compact objects and theoretical scenarios extending beyond the standard black hole paradigm. All metrics considered are static and spherically symmetric, and can be expressed in the general form
\begin{equation}
    ds^2 = -f(r)\,dt^2 + g(r)\,dr^2 + R(r)\,(d\theta^2 + \sin^2\theta\,d\phi^2) \, ,
    \label{eq:metric}
\end{equation}
where the metric functions $f(r)$, $g(r)$, and $R(r)$ encode the spacetime curvature properties associated with a given gravitational source. To investigate potential deviations from the Schwarzschild geometry, we consider a set of static, spherically symmetric spacetimes, which are discussed in this section. Deviations from the Schwarzschild solution are quantified through a generalized charge-like parameter $q$, expressed in units of the gravitational mass $M$. In all cases, the limit $q = 0$ smoothly recovers the Schwarzschild metric. By statistically comparing these spacetimes against observational data, we can identify models that provide the best fit to the S2 orbit and rule out those that are significantly disfavored. This approach allows direct constraints on the geometry of Sgr A* from stellar dynamics. When combined with independent bounds from black hole shadow measurements~\cite{Vagnozzi:2022moj}, our analysis offers a more comprehensive understanding of the nature of the Milky Way’s GC.

\subsection{Schwarzschild Spacetime}

The Schwarzschild solution describes the exterior gravitational field of a static, spherically symmetric, uncharged mass. In spherical coordinates, the line element is
\begin{eqnarray}
    f(r) &= &\left(1-\frac{2M}{r}\right), \\
        g(r)&= &\left(1-\frac{2M}{r}\right)^{-1}, \\ 
        R(r)&=&r^2,
        \label{rn_metric}
\end{eqnarray}
where $M$ is the total mass. The spacetime has a curvature singularity at $r=0$, hidden behind the event horizon at $r=2M$. Its physical relevance is established by the Oppenheimer–Snyder–Datt (OSD) model~\cite{Oppenheimer:1939ue,datt}, which shows that a homogeneous, pressureless dust cloud collapses to form a Schwarzschild black hole. Birkhoff’s theorem further ensures that any spherically symmetric exterior spacetime must be Schwarzschild, with the event horizon forming dynamically to prevent matter or radiation from escaping~\cite{Joshi:2011rlc}.

\subsection{Reissner-Nordström Spacetime}
 The RN spacetime is the simplest generalization of the Schwarzschild geometry that incorporates an electric charge $q$~\cite{Reissner1916,Jeffery:1921}. It is a static, spherically symmetric, and asymptotically flat solution of the Einstein-Maxwell equations, describing the gravitational field of a non-rotating charged mass. The line element takes the form
\begin{eqnarray}
    f(r) &= &\left(1 - \frac{2M}{r} + \frac{q^2}{r^2}\right), \\
        g(r)&= &\left(1 - \frac{2M}{r} + \frac{q^2}{r^2}\right)^{-1}, \\ 
        R(r)&=&r^2,
        \label{rn_metric}
\end{eqnarray}
where $q$ denotes the electric charge of the compact object. The horizons of the RN spacetime are determined by the condition $f(r) = 0$ (or $g(r) = 0$ for static geometries), corresponding to coordinate singularities of the metric. Solving this condition yields
\begin{equation}
1 - \frac{2M}{r} + \frac{q^2}{r^2} = 0 \, ,
\label{rn_horizon1}
\end{equation}
whose solutions are
\begin{equation}
r_{\pm} = M \pm \sqrt{M^2 - q^2} \, .
\label{rn_horizon2}
\end{equation}
The outer radius $r_+$ corresponds to the event horizon, while $r_-$ denotes the Cauchy (inner) horizon. For $q/M < 1$, both horizons exist and the spacetime describes a non-extremal charged black hole. The extremal configuration occurs for $q/M = 1$, where the two horizons coincide at $r = M$. When $q/M > 1$, no real roots exist, and the central singularity at $r = 0$ becomes naked, smoothly connecting the RN geometry to the class of naked singularity spacetimes. Thus, the motivation for studying the RN family is that it serves as a natural link between black holes and horizonless compact objects.

\subsection{Bardeen Spacetime}
The Bardeen black hole represents one of the earliest examples of a regular (non-singular) solution in general relativity~\cite{Bardeen:2018frm,AyonBeato:2000}. It is a static, spherically symmetric, and asymptotically flat spacetime that can be interpreted as the gravitational field of a self-gravitating magnetic monopole, sourced by a specific nonlinear electrodynamics (NED) Lagrangian~\cite{AyonBeato:2005}. The line element can be expressed through the metric functions
\begin{eqnarray}
    f(r) &=& 1 - \frac{2 M r^2}{\left(r^2 + q^2\right)^{3/2}}, \\
    g(r) &=& \left(1 - \frac{2 M r^2}{\left(r^2 + q^2\right)^{3/2}}\right)^{-1}, \\
    R(r) &=& r^2,
    \label{bd_metric}
\end{eqnarray}
where $M$ is the gravitational mass of the object and $q$ is the magnetic charge associated with the NED source. The function $f(r)$ ensures that the spacetime remains regular at the origin, where it behaves like a de Sitter core rather than a curvature singularity. Specifically, in the limit $r \rightarrow 0$, the metric approaches
\begin{equation}
    f(r) \approx 1 - \frac{2M}{q^3} r^2,
\end{equation}
which mimics a de Sitter geometry characterized by an effective cosmological constant $\Lambda_{\mathrm{eff}} = 6M/q^3$. At large radii ($r \gg q$), the metric smoothly reduces to the Schwarzschild form, recovering the expected asymptotic behavior. The horizons of the Bardeen spacetime are obtained by solving the condition $f(r) = 0$, i.e.,
\begin{equation}
    1 - \frac{2 M r_h^2}{\left(r_h^2 + q^2\right)^{3/2}} = 0 \, .
    \label{bd_horizon1}
\end{equation}
This equation admits two distinct real roots for regular black hole
\begin{equation}
\frac{q}{M} \leq \sqrt{\frac{16}{27}} \approx 0.77,
\end{equation}
corresponding to the event and Cauchy horizons. The extremal regular black hole configuration occurs at
\begin{equation}
\frac{q}{M} = \sqrt{\frac{16}{27}},
\end{equation}
where the two horizons coincide. For larger values,
\begin{equation}
\frac{q}{M} > \sqrt{\frac{16}{27}},
\end{equation}
no horizon forms and the solution describes a globally regular horizonless compact object. Unlike the RN black hole, the Bardeen geometry avoids curvature singularities at $r=0$, making it an appealing candidate for a quantum-corrected or NED-sourced compact object. The regularity of curvature invariants such as the Ricci and Kretschmann scalars supports its interpretation as a physically consistent, singularity-free spacetime. Thus, the motivation for studying the Bardeen black hole is that it offers a theoretical framework that bridges classical black holes and possible quantum gravity corrections at small scales.

\subsection{Hayward Spacetime}
The Hayward black hole is another prominent example of a regular, static, and spherically symmetric spacetime that eliminates the central singularity of the Schwarzschild geometry~\cite{Hayward:2005gi}. Like the Bardeen model, it introduces a de Sitter-like core near the origin, ensuring that all curvature invariants remain finite. The line element can be expressed in the general form
\begin{eqnarray}
    f(r) &=& 1 - \frac{2 M r^2}{r^3 + 2 q^2 M}, \\
    g(r) &=& \left(1 - \frac{2 M r^2}{r^3 + 2 q^2 M}\right)^{-1}, \\
    R(r) &=& r^2,
    \label{hy_metric}
\end{eqnarray}
where $M$ is the total mass of the compact object and $q$ is a length-scale parameter, often associated with the fundamental quantum-gravity cutoff or a characteristic curvature radius~\cite{Addazi:2022}. At large distances ($r \gg q$), the metric approaches the Schwarzschild limit,
\begin{equation}
    f(r) \approx 1 - \frac{2M}{r} + \mathcal{O}(r^{-4}),
\end{equation}
demonstrating asymptotic flatness and recovering the standard general relativistic behavior. In the near core region ($r \rightarrow 0$), the lapse function behaves as
\begin{equation}
    f(r) \approx 1 - \frac{r^2}{q^2},
\end{equation}
indicating that the geometry approaches a de Sitter spacetime with an effective cosmological constant $\Lambda_{\mathrm{eff}} = 3/q^2$. This feature guarantees that the curvature invariants, such as the Ricci scalar and Kretschmann scalar, remain finite everywhere. The horizon structure is determined by the condition $f(r)=0$, which yields
\begin{equation}
1 - \frac{2 M r_h^2}{r_h^3 + 2 q^2 M} = 0 \, .
\label{hy_horizon1}
\end{equation}
This equation admits two real positive roots for
$q/M < \sqrt{16/27} \approx 0.77$,
corresponding to the event and Cauchy horizons. The extremal configuration occurs at
$q/M = \sqrt{16/27}$,
where the two horizons coincide. For larger values of $q/M$, no horizons exist and the spacetime describes a globally regular horizonless compact object.

Physically, the parameter $q$ introduces a natural cutoff scale that prevents the divergence of curvature invariants, often interpreted as an effective manifestation of quantum gravitational effects or an upper bound on the matter density~\cite{Vagnozzi:2022moj}. The Hayward metric therefore provides a phenomenologically motivated framework for exploring singularity resolution and quantum-corrected black hole spacetimes. For suitable ranges of the parameter $q$, the solution satisfies the weak energy condition, reinforcing its consistency as an effective classical model arising from an underlying nonlinear electrodynamics or modified gravity source.

\subsection{Simpson-Visser Spacetime}  
The SV spacetime represents a one-parameter regular extension of the Schwarzschild geometry that interpolates smoothly between black holes and traversable wormholes through the introduction of a regularization parameter $q > 0$~\cite{Simpson:2018tsi,Bambhaniya:2021ugr}. It is static, spherically symmetric, and asymptotically flat, with the line element given by  
\begin{eqnarray}
    f(r) &=& \left(1 - \frac{2M}{\sqrt{r^2 + q^2}}\right), \\
    g(r) &=& \left(1 - \frac{2M}{\sqrt{r^2 + q^2}}\right)^{-1}, \\
    R(r) &=& r^2 + q^2, 
    \label{sv_metric}
\end{eqnarray}
where $M$ is the Arnowitt-Deser-Misner (ADM) mass of the spacetime and $q$ is a positive parameter that removes the central singularity present in the Schwarzschild solution. The horizons of the SV spacetime are determined from the condition $f(r) = 0$, which leads to
\begin{equation}
1 - \frac{2M}{\sqrt{r^2 + q^2}} = 0 \, ,
\label{sv_horizon1}
\end{equation}
whose real roots are given by
\begin{equation}
r_{\pm} = \pm \sqrt{(2M)^2 - q^2} \, .
\label{sv_horizon2}
\end{equation}
The nature of the geometry depends on the value of $q$ relative to $2M$:
\begin{enumerate}
    \item $q = 0$: Schwarzschild black hole with a curvature singularity at $r = 0$;
    \item $0 < q < 2M$: regular black hole with two horizons at $r = \pm \sqrt{(2M)^2 - q^2}$;
    \item $q = 2M$: extremal configuration with a null throat at $r = 0$;
    \item $q > 2M$: two-way traversable wormhole with a timelike throat at $r = 0$.
\end{enumerate}
At large distances ($r \gg q$), the function $f(r)$ approaches
\begin{equation}
f(r) \approx 1 - \frac{2M}{r} + \frac{Mq^2}{r^3} + \mathcal{O}\!\left(\frac{1}{r^5}\right),
\end{equation}
demonstrating that the SV spacetime is asymptotically Schwarzschild and hence asymptotically flat.  

The parameter $q$ regularizes the central region, ensuring that all polynomial curvature invariants remain finite for $q > 0$, and replaces the singularity with a smooth throat of finite areal radius $R(0) = q$. This property allows the SV spacetime to serve as a physically consistent bridge between regular black holes and traversable wormholes, offering a useful model for investigating horizonless compact objects~\cite{Arora:2023ltv}. The geometric structure of the spacetime in the regime where it represents a regular black hole is rather unconventional. Instead of exhibiting a bounce back into the same universe, the spacetime describes a transition into a future branch of the universe~\cite{Barcelo:2014cla,BenAchour:2020gon}. As a result, this geometry does not fall into the class of standard regular black hole solutions such as the Bardeen, Bergmann-Roman, Frolov, or Hayward metrics~\cite{Hayward:2005gi,Frolov:2017rjz,Cano:2018aod,Bardeen:2018frm}. 

Recent investigations have highlighted that several widely studied regular black hole models, including the analytically extended Hayward black hole and the non-analytic smooth black hole proposed by Culetu, Simpson, and Visser, suffer from geodesic incompleteness~\cite{Zhou:2022yio}. A similar analysis can therefore be carried out for the SV metric to assess its geodesic completeness~\cite{Arora:2023ltv}. Scalar curvature invariants provide a powerful, coordinate-independent method to characterize the presence of horizons in the SV spacetime~\cite{Abdel,Coley:2017woz,mcnut}. In particular, one may construct the following dimensionless scalar invariant~\cite{Abdel},
\begin{eqnarray}
\mathcal{Q}_2=\frac{1}{27}\frac{I_5 I_6 - I_7^2}{\left(I_1^2 + I_2^2\right)^{5/2}},
\label{5}
\end{eqnarray}
where $I_5 = k_\mu k^\mu$ with $k_\mu = -\nabla_\mu I_1$, $I_6 = n_\mu n^\mu$ with $n_\mu = -\nabla_\mu I_2$, and $I_7 = k_\mu n^\mu$. Here, $I_1 = C_{abcd}C^{abcd}$ and $I_2 = C^*_{abcd}C^{abcd}$ denote the quadratic Weyl invariant and its dual, respectively. For $q<2M$, the invariant $\mathcal{Q}_2$ vanishes at $r=0$ as well as at $r=\sqrt{(2M)^2-q^2}$, corresponding to the event horizon. This behavior holds for all angular coordinates, except on the equatorial plane $\theta=\pi/2$, where $\mathcal{Q}_2$ identically vanishes for all $r$. In contrast, for $q>2M$, $\mathcal{Q}_2$ vanishes only at $r=0$, indicating the absence of an event horizon. Consequently, in the SV spacetime, the condition $\mathcal{Q}_2=0$ for $r>0$ and $\theta\neq\pi/2$ is consistent with the presence of an event horizon, understood as a null hypersurface $\Sigma=\partial J^-(\mathcal{I}^+)\cap\mathcal{M}$, where $J^-(\mathcal{I}^+)$ denotes the causal past of future null infinity. Importantly, this characterization remains valid in any coordinate system~\cite{Arora:2023ltv}.

An additional essential property of the spacetime is the finiteness of its ADM mass, which is closely related to asymptotic flatness. While asymptotically Minkowskian spacetimes typically possess finite ADM mass, there exist quasi-flat spacetimes that fail to do so~\cite{Sud}. The ADM mass for a spacetime manifold $(\mathcal{M}, g_{\alpha\beta})$ is defined as
\begin{equation}
M_{\rm ADM} = -\frac{1}{8\pi}\lim_{S_t\to\infty}\oint_{S_t}(K-K_0)\sqrt{b}\,d^2y,
\label{6}
\end{equation}
where $S_t$ is a closed spacelike two-surface, $K$ is its extrinsic curvature when embedded in the spacelike hypersurface $\Sigma_t$, and $K_0$ is the corresponding extrinsic curvature in flat space. The surface element is given by $\sqrt{b}\,d^2y=(r^2+q^2)\sin\theta\,d\theta\,d\phi$, with $b$ denoting the determinant of the induced metric on $S_t$. For the SV spacetime, the extrinsic curvature of $S_t$ takes the form
\begin{eqnarray}
K=\frac{2r}{r^2+q^2}\left(1-\frac{2M}{\sqrt{r^2+q^2}}\right)^{1/2},
\label{7}
\end{eqnarray}
while $K_0=2/r$. Substituting these expressions into Eq.~(\ref{6}), one finds
\[
M_{\rm ADM}
= \lim_{r\to\infty}\left[r\left(1+\frac{q^2}{r^2}\right)
- r\sqrt{1-\frac{2M}{\sqrt{r^2+q^2}}}\right]
= M,
\]
which is finite. Moreover, the spacetime is asymptotically Minkowskian, as evidenced by the limits $\lim_{r\to\infty} g_{tt}=-1$, $\lim_{r\to\infty} g_{rr}=1$, $\lim_{r\to\infty} g_{\theta\theta}=r^2$, and $\lim_{r\to\infty} g_{\phi\phi}=r^2\sin^2\theta$. This analysis ensures that, the SV spacetime is asymptotically flat. Here we take metric functions as $f(r),g(r),R(r)$ instead of $g_{tt},g_{rr},g_{\phi\phi}$.

\subsection{Janis-Newman-Winicour Spacetime}
The JNW solution represents a static, spherically symmetric spacetime sourced by a massless scalar field minimally coupled to gravity~\cite{Janis1968,Virbhadra:1997}. It is an exact solution of the Einstein-Klein-Gordon equations and can be viewed as the most general scalar-field extension of the Schwarzschild geometry. The corresponding line element takes the form
\begin{eqnarray}
    f(r) &=& \left(1 - \frac{b}{r}\right)^{\nu}, \\
    g(r) &=& \left(1 - \frac{b}{r}\right)^{-\nu}, \\
    R(r) &=& r^2\left(1-\frac{b}{r}\right)^{(1-\nu)},
    \label{jnw_metric}
\end{eqnarray}
where the ADM mass and the scalar charge
parameters are related through
\begin{equation}
    b = 2\sqrt{M^2 + q^2}, \qquad 
    \nu = \frac{2M}{b} = \frac{1}{\sqrt{1 + (q/M)^2}} \, .
\end{equation}
The spacetime exhibits a curvature singularity at $r = b$~\cite{Bambhaniya2024}, which is not shielded by any event horizon. Therefore, for any nonzero scalar charge ($q \neq 0$), the singularity is horizonless. In the limit $q \rightarrow 0$ (or equivalently $\nu \rightarrow 1$), the solution smoothly reduces to the Schwarzschild metric, restoring the standard black hole geometry. A distinctive feature of the JNW spacetime is that the coordinate $r$ does not coincide with the areal radius, since $R^2(r) \neq r^2$. The true areal radius is determined by
\begin{equation}
    R_{\mathrm{areal}}(r) = \sqrt{R(r)} = r\left(1-\frac{b}{r}\right)^{(1-\nu)/2},
\end{equation}
which introduces significant geometric deviations from Schwarzschild near the central region. The parameter $\nu$ controls the degree of deviation, smaller $\nu$ (or larger scalar charge) leads to stronger curvature and more prominent naked singularity features.

Physically, the JNW geometry interpolates between flat space and the Schwarzschild black hole through the scalar charge $q$. It provides a minimal model of a scalar field supported compact object without horizons. Its analytical simplicity makes it a widely used benchmark for testing the cosmic censorship conjecture~\cite{Penrose:1969,Joshi:2011rlc}. It also provides a framework to explore observational differences between naked singularities and black holes~\cite{Bambhaniya2024b,Patel:2023efv,Kalsariya:2024qyp,Bambhaniya:2025qoe,Acharya:2023vlv}. These investigations reveal characteristic signatures that could help identify horizonless compact objects in astrophysical observations.

\section{Energy Conditions}
\label{sec:3}

Since Schwarzschild geometry is a vacuum solution of Einstein’s equations, the associated energy-momentum tensor vanishes identically, implying
\begin{equation}
\rho = 0, 
\qquad 
p_r = p_\theta = p_\phi = 0 ,
\end{equation}
where $\rho$ denotes the energy density, $p_r$ the radial pressure, and $p_\theta$, $p_\phi$ the tangential pressures. As a consequence, the standard energy conditions are trivially satisfied. The RN and JNW spacetimes also satisfy the standard energy conditions within classical general relativity~\cite{Bambhaniya2019a}. For the SV, Bardeen, and Hayward spacetimes, the geometry may be interpreted in terms of an effective stress-energy tensor obtained from the Einstein equations. In contrast to the Schwarzschild case, these geometries are supported by nonvacuum effective matter sources associated with the regularization of the central singularity.

For the SV spacetime, the effective stress-energy tensor components outside the horizon are given by~\cite{Simpson:2018tsi}
\begin{equation}
\rho_{\rm SV}(r)
=
-\frac{q^2\left(\sqrt{r^2+q^2}-4M\right)}
{8\pi (r^2+q^2)^{5/2}},
\end{equation}
\begin{equation}
p_{r,{\rm SV}}(r)
=
-\frac{q^2}
{8\pi (r^2+q^2)^2},
\end{equation}
\begin{equation}
p_{t,{\rm SV}}(r)
=
\frac{q^2\left(\sqrt{r^2+q^2}-M\right)}
{8\pi (r^2+q^2)^{5/2}},
\end{equation}
where $p_t=p_\theta=p_\phi$ denotes the tangential pressure. All these quantities remain finite throughout the spacetime, including at $r=0$, confirming the regular nature of the geometry. The null energy condition (NEC) requires
\begin{equation}
\rho+p_r \geq 0,
\qquad
\rho+p_t \geq 0.
\end{equation}
For the SV spacetime one obtains
\begin{equation}
\rho_{\rm SV}+p_{r,{\rm SV}}
=
-\frac{
q^2\left|\sqrt{r^2+q^2}-2M\right|
}{
4\pi (r^2+q^2)^{5/2}
},
\end{equation}
which is negative everywhere except possibly on the horizon~\cite{Simpson:2018tsi}. Therefore, the NEC is violated throughout the spacetime. Consequently, the weak, dominant, and strong energy conditions are also violated.

For the Bardeen spacetime, the geometry may be interpreted as a solution of Einstein gravity coupled to nonlinear electrodynamics \cite{Ayon-Beato:2000mjt}. The corresponding effective stress-energy tensor components obtained from the Einstein equations are
\begin{equation}
\rho_{\rm BD}(r)
=
\frac{3M q^2}
{4\pi (r^2+q^2)^{5/2}},
\end{equation}
\begin{equation}
p_{r,{\rm BD}}(r)
=
-\rho_{\rm BD}(r),
\end{equation}
\begin{equation}
p_{t,{\rm BD}}(r)
=
\frac{3M q^2(3r^2-2q^2)}
{8\pi (r^2+q^2)^{7/2}}.
\end{equation}
all effective quantities remain finite throughout the spacetime, including at $r=0$. The NEC for the Bardeen spacetime
\begin{equation}
\rho_{\rm BD}+p_{r,{\rm BD}} = 0,
\end{equation}
and
\begin{equation}
\rho_{\rm BD}+p_{t,{\rm BD}}
=
\frac{15M q^2 r^2}
{8\pi (r^2+q^2)^{7/2}}
\geq 0,
\end{equation}
therefore, the NEC is satisfied everywhere, saturating in the radial direction. Since
\begin{equation}
\rho_{\rm BD}(r)\geq0
\end{equation}
for all $r$, the weak energy condition is also satisfied everywhere. The strong energy condition requires
\begin{equation}
\rho+p_r+2p_t \geq 0,
\end{equation}
using the above expressions,
\begin{equation}
\rho_{\rm BD}+p_{r,{\rm BD}}+2p_{t,{\rm BD}}
=
\frac{3M q^2(3r^2-2q^2)}
{4\pi (r^2+q^2)^{7/2}},
\end{equation}
this quantity becomes negative for
\begin{equation}
r^2 < \frac{2q^2}{3},
\end{equation}
indicating a violation of the strong energy condition near the regular core \cite{Ayon-Beato:2000mjt}.

For the Hayward spacetime,
the effective stress-energy tensor components obtained from the Einstein equations are \cite{Hayward:2005gi}
\begin{equation}
\rho_{\rm HY}(r)
=
\frac{3M^{2}q^{2}}
{2\pi (r^{3}+2Mq^{2})^{2}},
\end{equation}
\begin{equation}
p_{r,{\rm HY}}(r)
=
-\rho_{\rm HY}(r),
\end{equation}
\begin{equation}
p_{t,{\rm HY}}(r)
=
\frac{3M^{2}q^{2}(r^{3}-Mq^{2})}
{\pi (r^{3}+2Mq^{2})^{3}},
\end{equation}
all effective quantities remain finite throughout the spacetime, including at $r=0$. The NEC for the Hayward spacetime
\begin{equation}
\rho_{\rm HY}+p_{t,{\rm HY}}
=
\frac{9M^{2}q^{2}r^{3}}
{2\pi (r^{3}+2Mq^{2})^{3}}
\geq 0,
\end{equation}
therefore, the NEC is satisfied everywhere, saturating in the radial direction. Since
\begin{equation}
\rho_{\rm HY}(r)\geq0
\end{equation}
for all $r$, the weak energy condition is also satisfied everywhere. The strong energy condition for Hayward spacetime is
\begin{equation}
\rho_{\rm HY}+p_{r,{\rm HY}}+2p_{t,{\rm HY}}
=
\frac{6M^{2}q^{2}(r^{3}-Mq^{2})}
{\pi (r^{3}+2Mq^{2})^{3}},
\end{equation}
this quantity becomes negative for
\begin{equation}
r^{3}<Mq^{2},
\end{equation}
indicating a violation of the strong energy condition near the regular core \cite{Hayward:2005gi}.

From a physical perspective, violations of the classical energy conditions do not necessarily imply pathologies or inconsistencies. Rather, the effective stress-energy tensor should be interpreted as encoding quantum gravitational, semiclassical, or nonlocal effects that become relevant at high curvature scales~\cite{Frolov:2016pav}. Quantum fields in curved spacetime are known to violate classical energy conditions while still satisfying suitable averaged or quantum inequalities~\cite{Ford:1995pb,Visser:1999de}. In the present context, localized violations of the strong or null energy conditions near the regular core or throat provide the mechanism through which curvature singularities are replaced by smooth regular geometries while preserving asymptotic flatness and a finite ADM mass.

\section{Geodesic Equation and Initial Conditions}
\label{sec:4}
The motion of a test particle in a curved spacetime is governed by the geodesic equation
\begin{equation}
\ddot{x}^\mu + \Gamma^{\mu}_{\nu\sigma}\dot{x}^\nu \dot{x}^\sigma = 0,
\label{eq:geodesic}
\end{equation}
where an overdot denotes differentiation with respect to the affine parameter, which for timelike geodesics is taken to be the proper time $\tau$, and $\Gamma^{\mu}_{\nu\sigma}$ are the Christoffel symbols associated with the spacetime metric. This equation represents a system of four coupled second-order differential equations for the spacetime coordinates
\[
(t(\tau),\, r(\tau),\, \theta(\tau),\, \phi(\tau)).
\]
A unique solution to the geodesic equations is obtained once suitable initial conditions are specified at a reference proper time $\tau_0$ for both the position four-vector,
\[
(t(\tau_0),\, r(\tau_0),\, \theta(\tau_0),\, \phi(\tau_0)),
\]
and the corresponding four-velocity,
\[
(\dot t(\tau_0),\, \dot r(\tau_0),\, \dot \theta(\tau_0),\, \dot \phi(\tau_0)).
\]
For timelike geodesics, the four-velocity satisfies the normalization condition
\begin{equation}
\dot{x}_\mu \dot{x}^\mu = -1,
\end{equation}
which provides a constraint relating the components of the four-velocity. Since the spacetimes under consideration are all spherically symmetric, the particle motion can always be confined to a plane without loss of generality. We therefore choose the equatorial plane,
\begin{equation}
\theta = \frac{\pi}{2}, \qquad \dot{\theta} = 0,
\end{equation}
which simplifies the equations of motion while preserving the full physical content of the problem.

Furthermore, because the spacetime is stationary, the geodesic equations do not depend explicitly on the coordinate time $t$. As a result, the choice of the initial time is arbitrary. In practice, however, it is convenient to select a physically meaningful reference point. We choose the time of apocenter passage of the S2 star as the initial epoch, since the radial velocity vanishes at this point and the orbital configuration is particularly simple. The orbit is then integrated both forward and backward in time from this reference epoch. With this choice, the initial conditions are specified by fixing the radial coordinate to the apocenter distance, setting the azimuthal angle to $\phi = \pi$, and imposing a vanishing radial velocity. The remaining components of the four-velocity are determined using the normalization condition together with the conserved quantities of motion. Explicitly, the initial conditions are given by
\begin{align}
t(\tau_0) &= t_{\mathrm{ref}}, \\
r(\tau_0) &= a(1+e), \\
\phi(\tau_0) &= \pi, \\
\dot r(\tau_0) &= 0 ,
\label{eq:initial_cond}
\end{align}
where  the orbit $a$ the semi-major axis and $e$ the eccentricity. This choice fully specifies the initial state of the system and allows the geodesic equations to be integrated consistently in order to obtain the particle’s trajectory. Further, the Lagrangian of the test particle is defined as,
\begin{equation}
    2\tilde{\mathcal{L}} = g_{\mu\nu}\dot x^\mu \dot x^\nu,
\end{equation}
which is used to derive the conserved quantities. Due to the symmetry of the spacetime, there are two conserved quantities namely, conserved energy per unit rest mass ($\tilde{E}$) corresponding to temporal symmetry and conserved angular momentum per unit rest mass ($\tilde{L}$) corresponding to the spherical symmetry. They can be expressed using the Lagrangian as follows,
\begin{equation}
    \tilde E = - \frac{\partial \tilde{\mathcal{L}}}{\partial \dot t} ,
\end{equation}
\begin{equation}
    \tilde L =  \frac{\partial \tilde{\mathcal{L}}}{\partial \dot \phi} .
\end{equation}
One can then derive the relation between the radial coordinate, specific energy and the specific angular momentum as,
\begin{equation}
    \dot r^2 = \tilde E^2 - V_{\mathrm {eff}}(r), 
\end{equation}
where $V_{\mathrm{eff}}$ is the effective potential described in terms of specific angular momentum.

\section{Data \& Methodology}
\label{sec:5}

In this section, we describe the dataset used to determine the best-fit orbit for
different spacetime models and outline the procedure adopted to construct the
corresponding observables. This involves projecting the relativistic trajectory onto
the plane of the sky and incorporating non-negligible timing effects, such as the
R\o mer delay and relativistic corrections to the line-of-sight (LoS) velocity. We use
\texttt{PyGRO}~\cite{DellaMonica:2025npq} to integrate the fully relativistic equations
of motion and project the resulting trajectories onto the sky plane. Finally, we infer
the orbital parameters for each spacetime model using an MCMC analysis and compare
their performance against the Schwarzschild geometry using statistical model–selection
criteria.

\subsection{Data}

We use the S2 star dataset presented by Gillessen et al.\ (2017)~\cite{Gillessen:2017jxc}. The dataset comprises 145 astrometric measurements in right ascension and declination spanning the period from 1992 to 2016, together with 44 radial-velocity measurements obtained between 2003 and 2016. This long observational baseline covers more than one full orbital period of S2, while the inclusion of radial-velocity data provides the essential line-of-sight information required to tightly constrain its highly eccentric orbit. To tightly constrain our model parameters, we also include the 1PN precession of the S2 star as the observed relativistic precession factor~\cite{Abuter2020}
\begin{equation}
    f_{\rm sp} = 1.10 \pm 0.19.
    \label{eq:obs_fsp}
\end{equation}

\subsection{Orbital Parameters}
Since the spacetimes considered in this work generally possess nontrivial angular sectors, the orbital plane coordinates must be constructed using the geometric radius associated with the angular part of the metric. For the general static and spherically symmetric line element as given in Eq. (\ref{eq:metric}),
the function $R(r)$ determines the physical angular geometry of the spacetime. Accordingly, the Cartesian coordinates of the orbit in the equatorial plane are defined as
\begin{equation}
x = \sqrt{R(r)}\cos\phi, \qquad
y = \sqrt{R(r)}\sin\phi, \qquad
z = 0.
\end{equation}
The corresponding coordinate-time velocity components are then given by
\begin{align}
v_x &= \frac{R'(r)}{2\sqrt{R(r)}}\,v_r\cos\phi
      - \sqrt{R(r)}\,v_\phi\sin\phi,\\
v_y &= \frac{R'(r)}{2\sqrt{R(r)}}\,v_r\sin\phi
      + \sqrt{R(r)}\,v_\phi\cos\phi,\\
v_z &= 0,
\end{align}
where $v_r = dr/dt$, $v_\phi = d\phi/dt$, and $R'(r)=dR/dr$. For the Schwarzschild spacetime, $R(r)=r^2$, recovering the standard relations $x=r\cos\phi$ and $y=r\sin\phi$. The remaining projection equations, and all later equations remain valid because they act on the already-defined physical Cartesian coordinates. The
corresponding four-velocity components follow from
$\dot r = v_r \dot x^0$ and $\dot \phi = v_\phi \dot x^0$. To compare the theoretical orbit with astrometric measurements, the model trajectory
must be projected onto the apparent sky plane. The transformation from the true orbital
frame $(x,y,z)$ to the observable frame $(X,Y,Z)$ is performed using the standard
$3\!-\!1\!-\!3$ Euler rotation,
\begin{equation}
X = xB + yG, \quad 
Y = xA + yF, \quad
Z = xC + yH,
\label{eq:sky_to_apparent_pos}
\end{equation}
and similarly for the velocities,
\begin{equation}
V_X = v_xB + v_yG, \quad 
V_Y = v_xA + v_yF, \quad
V_Z = v_xC + v_yH,
\label{eq:sky_to_apparent_velocity}
\end{equation}
where $(A,B,C,F,G,H)$ are the Thiele-Innes constants, which are determined by the orbital inclination $i$, the argument of pericenter $\omega$, and the longitude of the ascending node $\Omega$. The projection is therefore performed using the metric dependent geometric radius encoded in $R(r)$. The measured angular coordinates (relative right ascension and declination) are modeled as
\begin{equation}
\Delta\alpha(t_{\text{obs}}) = \frac{X}{D}
+ x_0 + v_{x0}(t_{\text{obs}} - t_{\text{ref}}),
\label{eq:ra}
\end{equation}
\begin{equation}
\Delta\delta(t_{\text{obs}}) = \frac{Y}{D}
+ y_0 + v_{y0}(t_{\text{obs}} - t_{\text{ref}}),
\label{eq:dec}
\end{equation}
where $D$ is the distance to the Milky Way GC, and $(x_0,y_0)$ and
$(v_{x0},v_{y0})$ are linear astrometric offset parameters required to compare the
theoretical model with the observations. The orbit may be parametrized either in terms of the conserved energy and angular momentum or, equivalently, using the semi-major axis and eccentricity. Since the latter provide a convenient and complete set of orbital parameters, we relate them to the radial turning points as
\begin{equation}
a = \frac{r_{\rm max}+r_{\rm min}}{2},
\end{equation}
\begin{equation}
e = \frac{r_{\rm max}-r_{\rm min}}{r_{\rm max}+r_{\rm min}},
\end{equation}
where $r_{\rm max}$ and $r_{\rm min}$ denote the apocenter and pericenter radii,
respectively. These turning points correspond to the roots of the radial equation
$\dot r = 0$, satisfying
\begin{equation}
V_{\rm eff}(r_{\rm min}) = V_{\rm eff}(r_{\rm max}) = \tilde E^2.
\end{equation}

To predict the apparent position of the star at a given epoch we include the R\o mer delay arising from the variation of the star’s LoS distance during its orbit. As a result, photons emitted at different epochs have different travel times. The difference between the emission time $t_{\rm em}$ and the observed time $t_{\rm obs}$ is given by~\cite{DamourDeruelle1986}
\begin{equation}
t_{\rm em} - t_{\rm obs} = \frac{Z(t_{\rm em})}{c}.
\label{eq:romer_delay}
\end{equation}
Due to this effect, photons emitted near apocenter arrive approximately one week later, while those emitted near pericenter arrive about half a day earlier (compared to the time they would take if they were emitted exactly at a distance $D$ from the observer). The Shapiro delay is neglected, as its contribution is at most $\sim 5$ minutes~\cite{Do2019} which is below the temporal resolution of the observations. Solving Eq.~\eqref{eq:romer_delay} provides a mapping between $t_{\rm em}$ and $t_{\rm obs}$. The observable LoS velocity includes both the longitudinal Doppler effect $\zeta_D$, arising from the projected kinematic velocity $V_Z$, and relativistic time dilation effects. The longitudinal Doppler contribution is
\begin{equation}
\zeta_D(t_{\text{obs}}) = \frac{V_Z(t_{\text{obs}})}{c}.
\label{doppler}
\end{equation}
In addition, the high orbital velocity and deep gravitational potential of S2 lead to gravitational redshift and transverse Doppler contributions, collectively referred to as the Einstein delay $\zeta_E$~\cite{Angelil2010,Gravity2018},
\begin{equation}
1 + \zeta_E(t_{\text{obs}}) = \frac{dt(t_{\text{obs}})}{d\tau}.
\label{einstein_delay}
\end{equation}
Combining these effects, the observed redshift can be expressed as an equivalent LoS velocity,
\begin{equation}
1 + \frac{v_{\rm LOS}(t_{\text{obs}})}{c} =
\left[1 + \zeta_D(t_{\text{obs}})\right]
\left[1 + \zeta_E(t_{\text{obs}})\right].
\end{equation}
This formulation follows the prescription of~\cite{DellaMonica:2025npq}. The full model
thus depends on the six orbital parameters
$(a,e,i,\omega,\Omega,t_{\rm peri})$, together with additional astrometric and spectroscopic nuisance parameters.

\begin{table}[h]
    \centering
    \small
    \renewcommand{\arraystretch}{1.2}
    \setlength{\tabcolsep}{16pt}
    \caption{Flat and Gaussian priors on all parameters}
    \begin{tabular}{|c|c|c|}
        \toprule
        \hline
        \hline
        \textbf{Flat Prior} 
        & \textbf{Start} 
        & \textbf{End} \\
        \midrule
        \hline
         $D$ (pc) & 5330 & 11330\\
         $ q \ (10^6 M_\odot)$ & 0 & 4.3 \\
         $M \ (10^6 M_\odot)$ & 3.700 & 4.900\\
         $t_p - 2018$ (yr) & -0.070 & 0.830 \\
         $a$ (mas) & 112 & 139 \\
         $e$ & 0.855 & 0.912\\
         $i$ (rad) & 2.237 & 2.447\\
         $\Omega$ (rad) &3.804 & 4.118\\
         $\omega$ (rad) & 0.994 & 1.292\\
         \hline
         \midrule 
         \textbf{Gaussian Prior} & \textbf{Mean} & $\sigma$ \\
         \hline
         $x_0$ (mas) & 0& 0.2\\
         $y_0$ (mas) & 0& 0.2\\
         $v_{x0}$ (mas $\rm yr^{-1}$) & 0& 0.1\\
         $v_{y0}$ (mas $\rm yr^{-1}$) & 0& 0.1\\
         $v_{\rm LSR}$ (km $\rm s^{-1}$) &0 & 5\\
        \bottomrule
        \hline
    \end{tabular}
    \label{tab:prior}
\end{table}

\subsection{MCMC Analysis}
\begin{figure*}[]
    \centering
    \subfloat[RN spacetime\label{fig:RN_constraint}]{\includegraphics[width=0.48\textwidth]{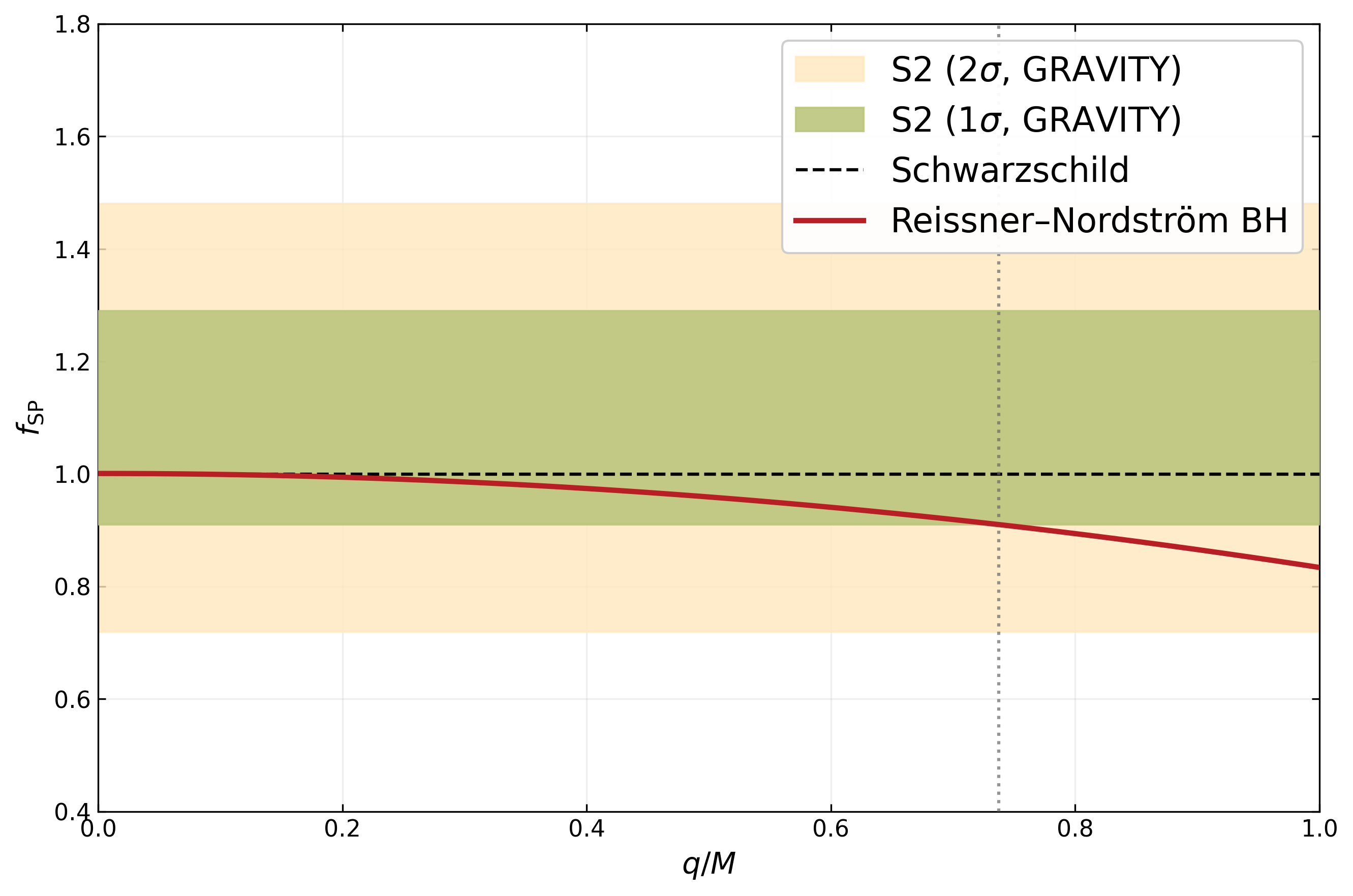}
    }\hfill
    \subfloat[Bardeen spacetime\label{fig:BD_constraint}]{\includegraphics[width=0.48\textwidth]{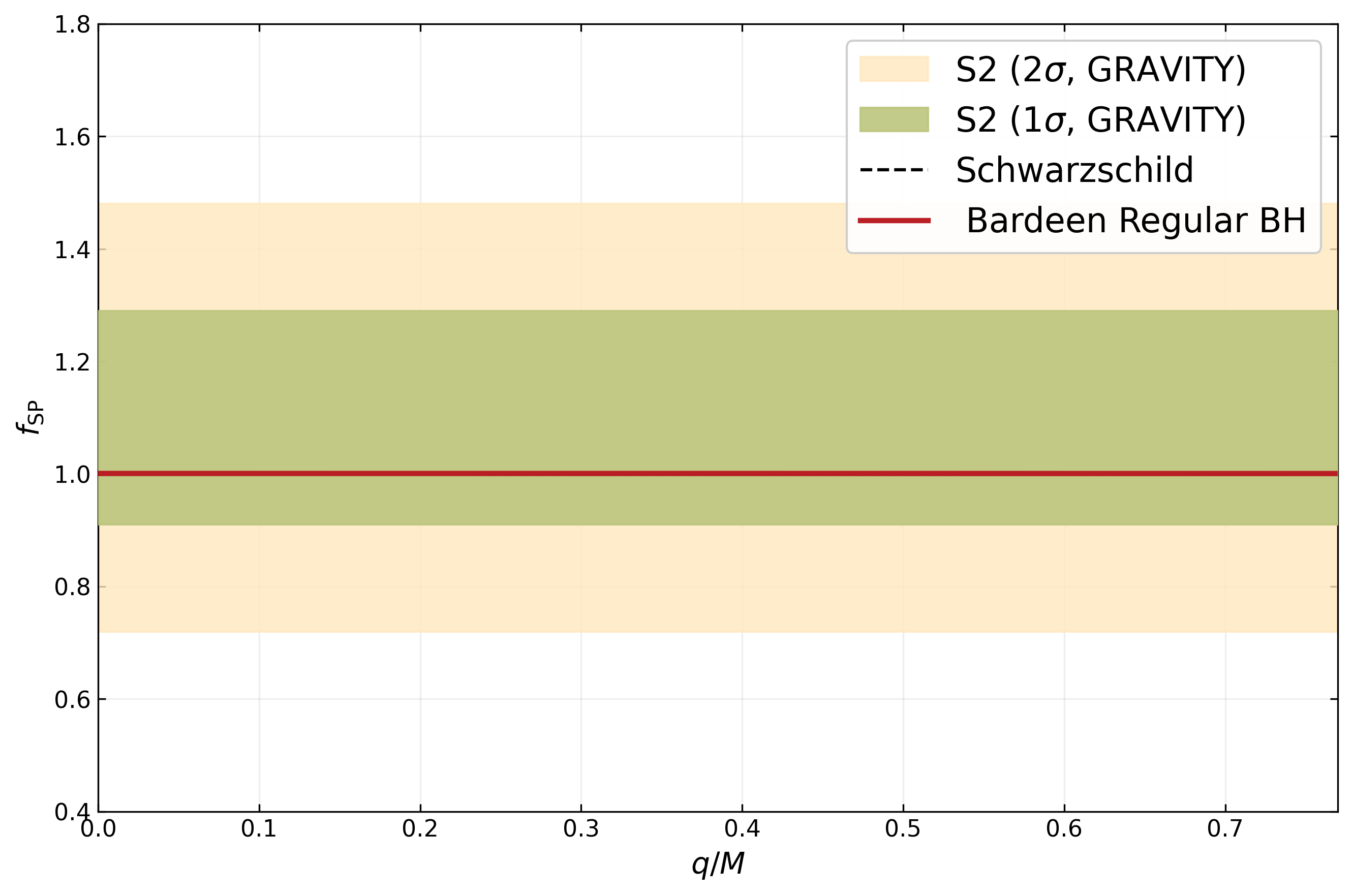}
    }\hfill
    \subfloat[Hayward spacetime\label{fig:HY_constraint}]{\includegraphics[width=0.48\textwidth]{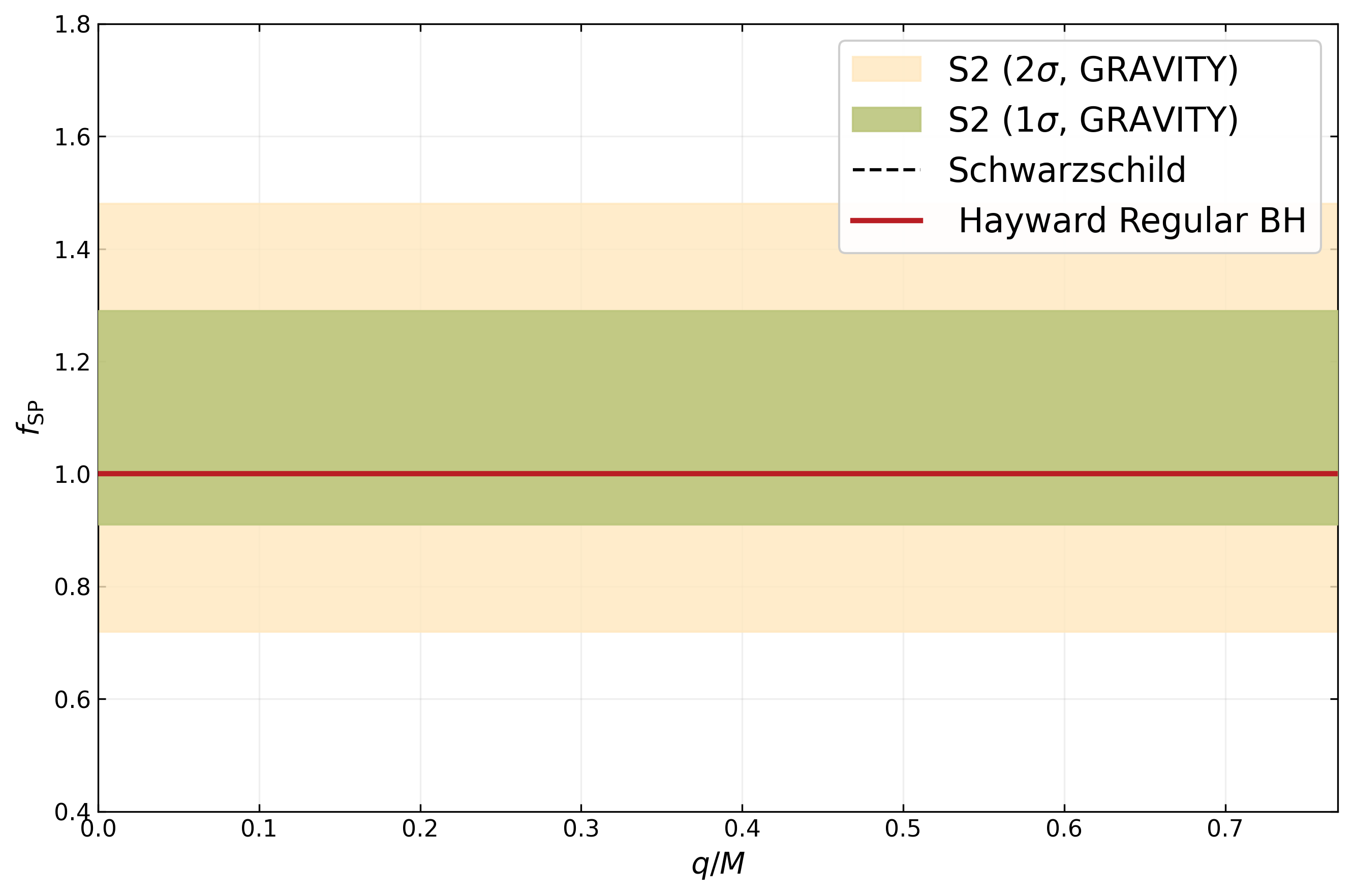}
    }
    \subfloat[SV spacetime\label{fig:SV_constraint}]{\includegraphics[width=0.48\textwidth]{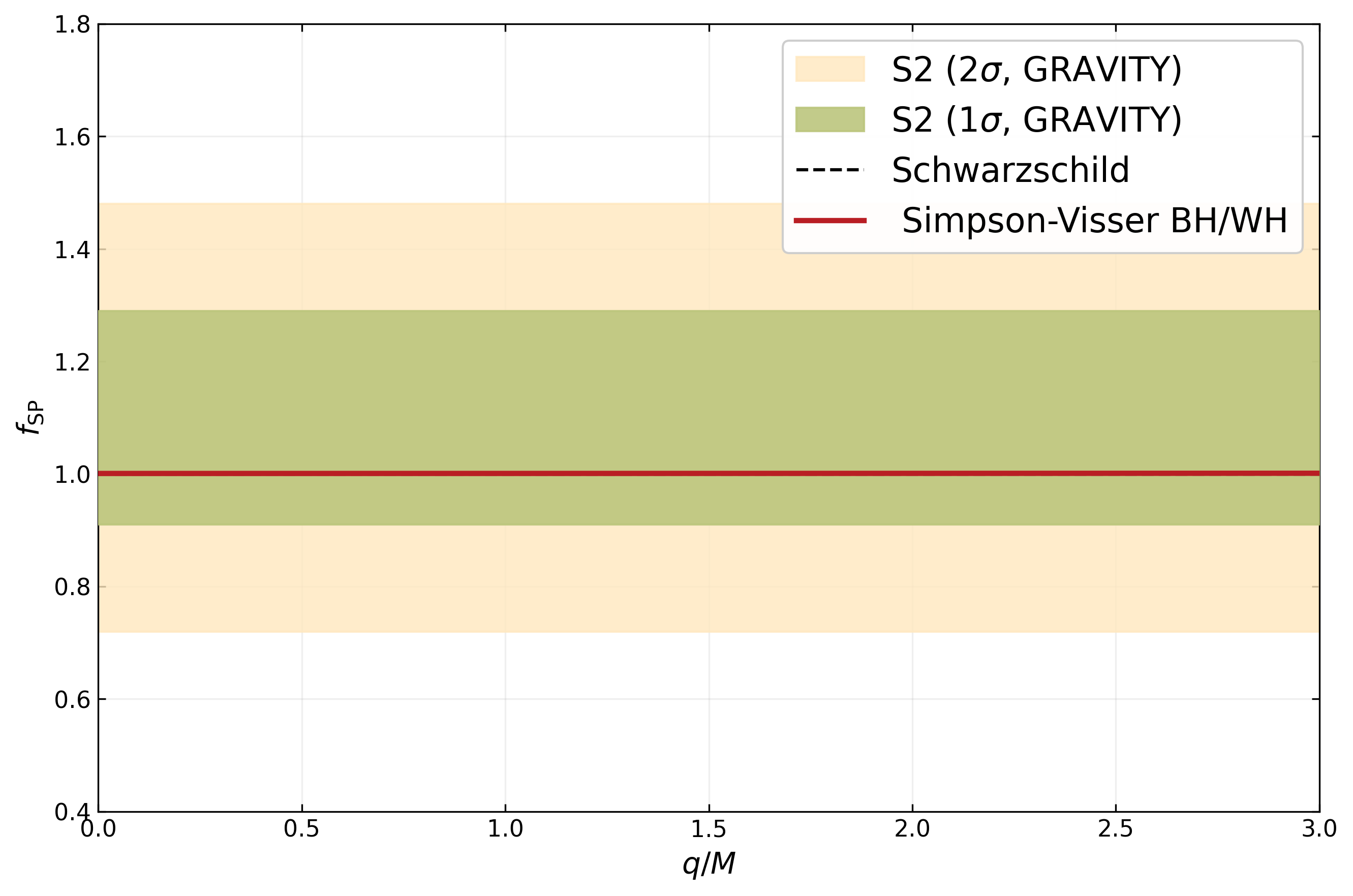}
    }\hfill
\subfloat[JNW spacetime\label{fig:JNW_constraint}]{\includegraphics[width=0.48\textwidth]{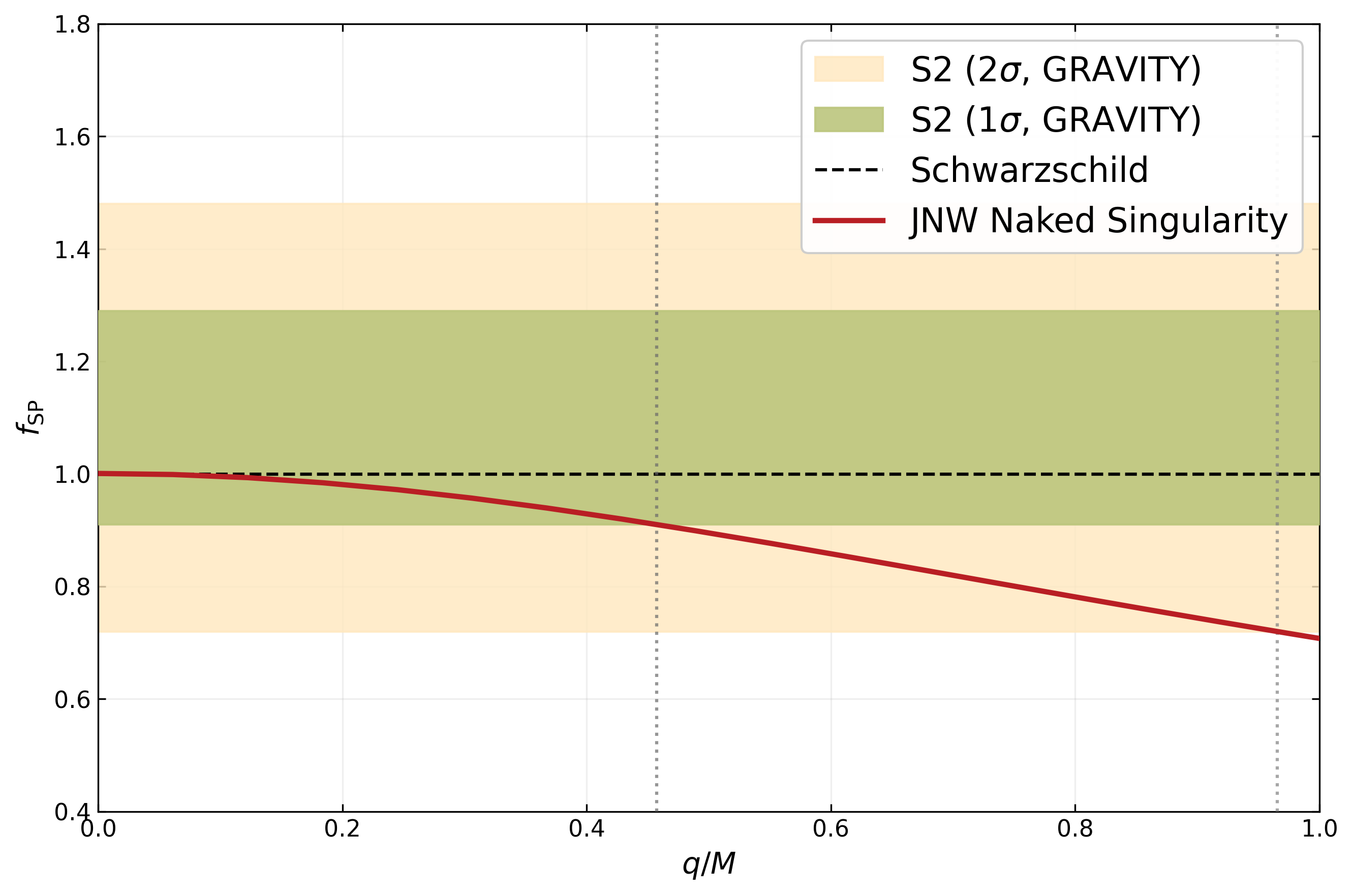}
}
\caption{Constraints on the generalized charge-like parameter $q$ inferred from the periastron precession of the S2 star. The shaded bands correspond to the $1\sigma$ and $2\sigma$ confidence intervals derived from the observed precession factor $f_{\rm sp}$. Each panel shows the allowed parameter range for a specific non-Schwarzschild spacetime model.}
\label{fig:constraints_all}
\end{figure*}
To infer the orbital parameters of the S2 star, we perform Bayesian parameter estimation
using Markov Chain Monte Carlo (MCMC) sampling. The complete parameter vector for all
spacetime models is
\[
\boldsymbol{\theta} =
\{ D, q, M, T_p, a, e, i, \Omega, \omega, x_0, y_0, v_{x0}, v_{y0}, v_{\rm LSR} \},
\]
forming a 14-dimensional parameter space. Here, $v_{\rm LSR}$ denotes the systemic LoS velocity offset accounting for a drift in the spectroscopic reference frame. For the Schwarzschild model, the charge-like parameter $q$
is absent, reducing the dimensionality to 13. The Bayesian framework provides a systematic probabilistic approach for estimating the model parameters $\boldsymbol{\theta}$ given a set of observational data
$\tilde{\mathcal D}$. Bayes' theorem relates the posterior distribution to the likelihood and the prior as
\begin{equation}
p(\boldsymbol{\theta} \mid \tilde{\mathcal D}) =
\frac{
\mathcal{L}(\tilde{\mathcal D} \mid \boldsymbol{\theta}) \,
\pi(\boldsymbol{\theta})
}{
\mathcal{Z}(\tilde{\mathcal D})
},
\end{equation}
where $\mathcal{L}(\tilde{\mathcal D} \mid \boldsymbol{\theta})$ is the likelihood
function, $\pi(\boldsymbol{\theta})$ is the prior distribution, and
$\mathcal{Z}(\tilde{\mathcal D})$ is the Bayesian evidence, which serves as a normalization constant. The observational
dataset consists of measurements
$\tilde{\mathcal D} = \{d^i\} \equiv \{ \alpha^i, \delta^i, v_{\rm RV}^i,f_{\rm sp} \}$ with associated
uncertainties $\sigma_d^i$, while the corresponding theoretical predictions are
$ \{ d^{\mathrm{th},i}\} \equiv
\{ \alpha^{\mathrm{th},i}(\boldsymbol{\theta}),
\delta^{\mathrm{th},i}(\boldsymbol{\theta}),
v_{\rm RV}^{\mathrm{th},i}(\boldsymbol{\theta}),f_{\rm sp}^{\rm th}(\boldsymbol{\theta}) \}$. Here $f_{\rm sp}^{\rm th}$ is calculated as, 
\begin{equation}
    f_{\rm sp }^{\rm th} = \frac{\Delta\phi_{\rm th}}{\Delta\phi_{\rm 1PN}}, 
    \label{eq:fsp}
\end{equation}
where $\Delta\phi_{\rm th}$ is calculated by integrating the geodesic motion and $\Delta\phi_{\rm 1PN}$ is calculated as, 
\begin{equation}
    \Delta\phi_{\rm 1PN} = \frac{6\pi M}{a(1-e^2)}.
    \label{eq:1PN}
\end{equation}

Assuming that the measurement errors are uncorrelated
\cite{Gillessen:2017jxc,Do2019}, the likelihood function is
\begin{equation}
\mathcal{L}(\tilde{\mathcal D} \mid \boldsymbol{\theta}) =
\prod_{i=1}^{N}
\exp\!\left[
-\frac{1}{2}
\left(
\frac{
d^i - d^{\mathrm{th},i}(\boldsymbol{\theta})
}{
\sigma_d^i
}
\right)^{\!2}
\right].
\label{eq:likelihood_general}
\end{equation}
Defining the chi-square statistic
\begin{equation}
\chi^2(\boldsymbol{\theta}) =
\sum_{i=1}^{N}
\left(
\frac{
d^i - d^{\mathrm{th},i}(\boldsymbol{\theta})
}{
\sigma_d^i
}
\right)^{\!2},
\label{eq:chi2_general}
\end{equation}
the likelihood can be written compactly as
\begin{equation}
\mathcal{L}(\tilde{\mathcal D} \mid \boldsymbol{\theta}) =
\exp\!\left[
-\frac{1}{2}\chi^2(\boldsymbol{\theta})
\right].
\label{eq:likelihood_chi2_general}
\end{equation}
Since the precession factor $f_{\rm sp}$ is not an independent observable but has been inferred from the same astrometric and spectroscopic dataset we use (plus additional data at and after the pericenter passage in 2018) ~\cite{Abuter2020}, we rescale the log-likelihood (\emph{i.e.} the $\chi^2$) by an overall factor of 2 to avoid effectively double-counting the same physical information as first done in Ref.~\cite{DellaMonica:2021xcf}.
The prior distribution encodes information about the parameters before incorporating observational data. We adopt a combination of flat (uniform) and Gaussian priors as given in the Table~\ref{tab:prior}. Flat priors are used for all parameters except the reference frame offsets, for which Gaussian priors are imposed. For a parameter with mean $\mu$ and standard deviation $\sigma$, the Gaussian prior is
\begin{equation}
\ln \pi(\theta) = -\frac{(\theta - \mu)^2}{2\sigma^2}.
\end{equation}
For the charge-like parameter $q$, the spacetime metric and hence the likelihood depend on $q^2$, implying a symmetry under $q \rightarrow -q$. To avoid sampling of two equivalent modes, we therefore reparameterize the model in terms of $q \geq 0$ and impose a flat prior on $q$. 
The total log-prior is given by the sum of the individual prior contributions. Posterior sampling is performed using the \texttt{emcee} ensemble sampler
\cite{ForemanMackey2013}, employing 32 walkers and 25\,000 steps (8,00,000 iterations). The first 3000 steps of each walker are discarded as burn-in. Convergence is verified using the Gelman-Rubin statistics, requiring $\hat{R} < 1.01$ for all sampled parameters.

To compare different spacetime models, we compute the Akaike Information Criterion (AIC)
and the Bayesian Information Criterion (BIC),
\begin{equation}
\mathrm{AIC} = \chi^2_{\rm min} + 2k,
\end{equation}
\begin{equation}
\mathrm{BIC} = \chi^2_{\rm min} + k \ln N,
\end{equation}
where $k$ is the number of free parameters, $N$ is the total number of data points, and
$\chi^2_{\rm min}$ is the minimum Chi-square value, i.e. the one computed with the best-fit combination of parameters. The strength of evidence relative to
the Schwarzschild model is interpreted as
\begin{itemize}
\item $0 \le \Delta \mathrm{AIC}, \Delta \mathrm{BIC} < 2$: statistically indistinguishable;
\item $2 \le \Delta \mathrm{AIC}, \Delta \mathrm{BIC} < 6$: weak evidence;
\item $6 \le \Delta \mathrm{AIC}, \Delta \mathrm{BIC} < 10$: strong evidence;
\item $\Delta \mathrm{AIC}, \Delta \mathrm{BIC} \ge 10$: very strong evidence.
\end{itemize}

\section{Constraints on the Charge-like Parameter from S2 Star Data}
\label{sec:constraint}

In this section, we derive observational constraints on the generalized charge-like parameter $q/M$ for the spacetime geometries introduced above by the measured pericenter precession of the S2 star orbiting Sgr~A*. The relativistic advance of the pericenter provides a clean and robust probe of the underlying spacetime geometry, as it directly reflects deviations in the effective gravitational potential from the Schwarzschild case. For each spacetime model, we compute the total per-orbit relativistic precession $\Delta\phi_{\rm th}$ by numerically integrating the timelike geodesic equations for bound stellar motion. To facilitate a direct comparison with observations, we normalize the theoretical precession to the first post-Newtonian (1PN) Schwarzschild prediction in Eq.~\eqref{eq:1PN} and thus, $f_{\rm sp}$ with Eq.~\eqref{eq:fsp}. In the Schwarzschild limit ($q/M \to 0$), one has $f_{\rm sp} \to 1$, while deviations from unity quantify non-Schwarzschild corrections to the spacetime geometry as shown in Figure~\ref{fig:constraints_all} for all models.

\begin{table}[h]
\centering
\small
\renewcommand{\arraystretch}{1.1}
\setlength{\tabcolsep}{4pt}
\caption{Constraints on generalized charge-like parameter $q/M$ from the S2 star observations.}
\label{tab:q_constraints}
\begin{tabular}{|c|cc|cc|}
\hline\hline
 & \multicolumn{2}{c|}{\textbf{MCMC Constraint}} 
 & \multicolumn{2}{c|}{\textbf{Analytical Constraint}}\\
 \hline
\textbf{Model}
 & $1\sigma$ & $2\sigma$
 & $1\sigma$ & $2\sigma$ \\
\hline
RN & $\leq  0.209$ & $\leq 0.307$ & $\leq0.738$ & $\leq1.000$ \\
BD & $\leq  0.642$ & $\leq 0.755$ & $\leq 0.770$ & $\leq 0.770$ \\
HY & $\leq0.644$ & $\leq0.759$ & $\leq 0.770$ & $\leq 0.770$ \\
SV & $\leq1.833$ & $\leq2.164$ & $<$ 2.000 & $<$ 2.000 \\
JNW & $\leq0.844$ & $\leq0.983$ & $\leq 0.457$ & $\leq 0.965$ \\
\hline
\end{tabular}
\end{table}

We confront the theoretical predictions for $f_{\rm sp}$ with the observational determination reported by the GRAVITY collaboration given in Eq.~\eqref{eq:obs_fsp}
and impose consistency at the $1\sigma$ and $2\sigma$ confidence levels. For each spacetime model, this procedure yields an allowed interval for the parameter $q/M$, corresponding to the range over which the predicted precession remains compatible with the observed value. As shown in Table~\ref{tab:q_constraints}, we report constraints on the generalized charge-like parameter $q/M$ for given spacetime models derived from S2 star observations. The MCMC-based constraints follow from Bayesian analysis as shown in Figure~\ref{fig:mcmc_constraint}, whereas the analytical bounds are obtained by comparison of the observed pericenter precession at the $1\sigma$ and $2\sigma$ confidence levels with the theoretical predictions as shown in Figures~\ref{fig:constraints_all}. These bounds provide direct, model-dependent constraints on departures from the Schwarzschild geometry in the strong-field regime probed by the S2 orbit. It is important to emphasize that the precession factor $f_{\rm sp}$ is not treated here as an independent observable in addition to the full astrometric and spectroscopic dataset. Rather, this section provides an analytic characterization of how the measured relativistic precession alone constrains the parameter $q/M$. As discussed in Sec.~\ref{sec:5}, the full Bayesian analysis accounts for correlations between orbital parameters and avoids double-counting of physical information. Therefore, the constraints derived here serve as a more qualitative benchmark that complements and aids the interpretation of the numerical MCMC-based parameter estimation.

\section{Results}
\label{sec:results}

\begin{figure}
    \centering
\includegraphics[width=0.96\linewidth]{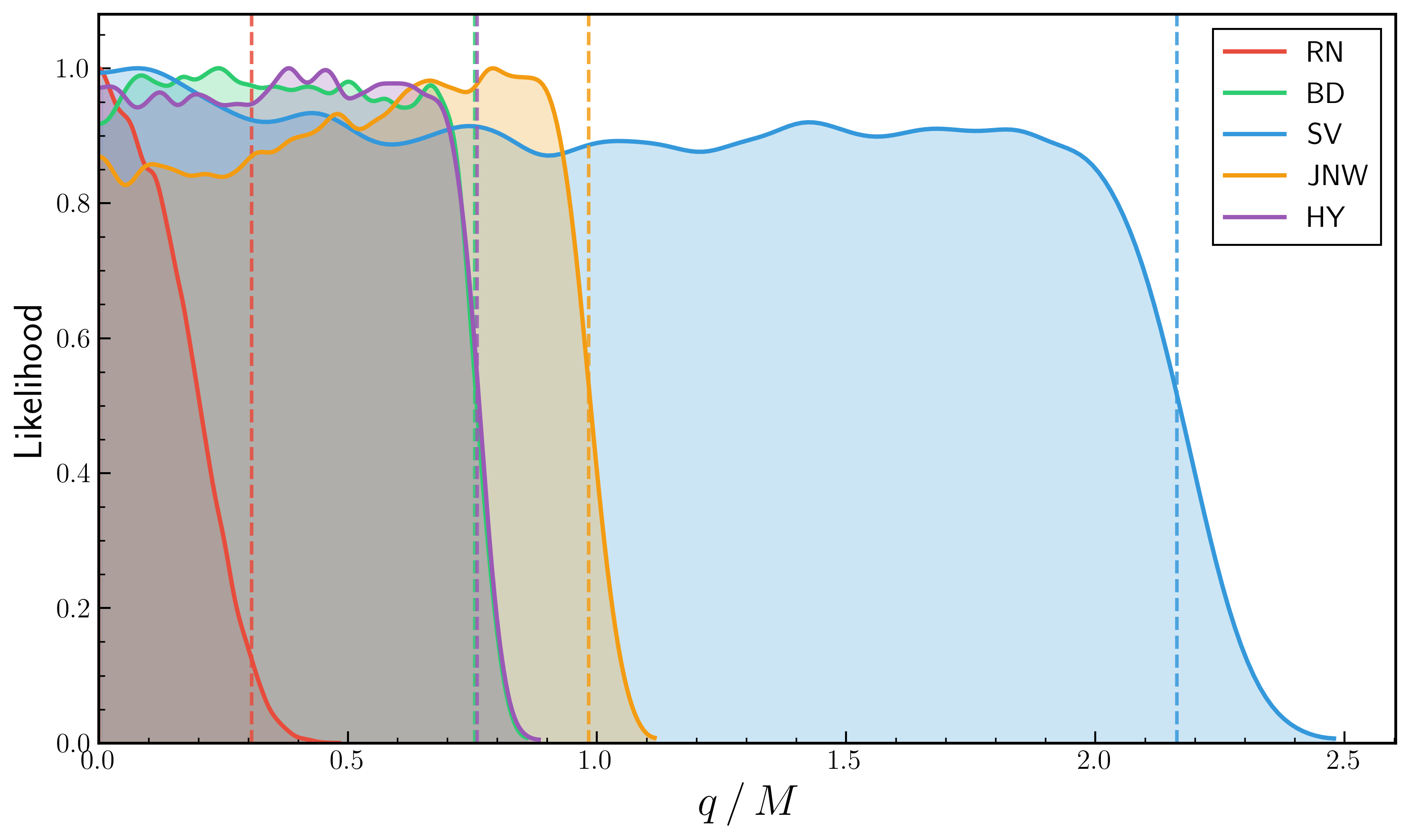}
\caption{Marginalized posterior distributions scaled to 1 for $q/M$ for all models obtained by MCMC analysis. The dotted line represents the 2$\sigma$ upper limit for $q/M$ for all spacetimes. The corresponding $1 \sigma$ and $2\sigma$ upper limits on $q/M$ are given in Table~\ref{tab:q_constraints}.}
\label{fig:mcmc_constraint}
\end{figure}

\begin{figure*}
    \subfloat[BD and SCH spacetime\label{BDfit}]{\includegraphics[width=0.48\textwidth]{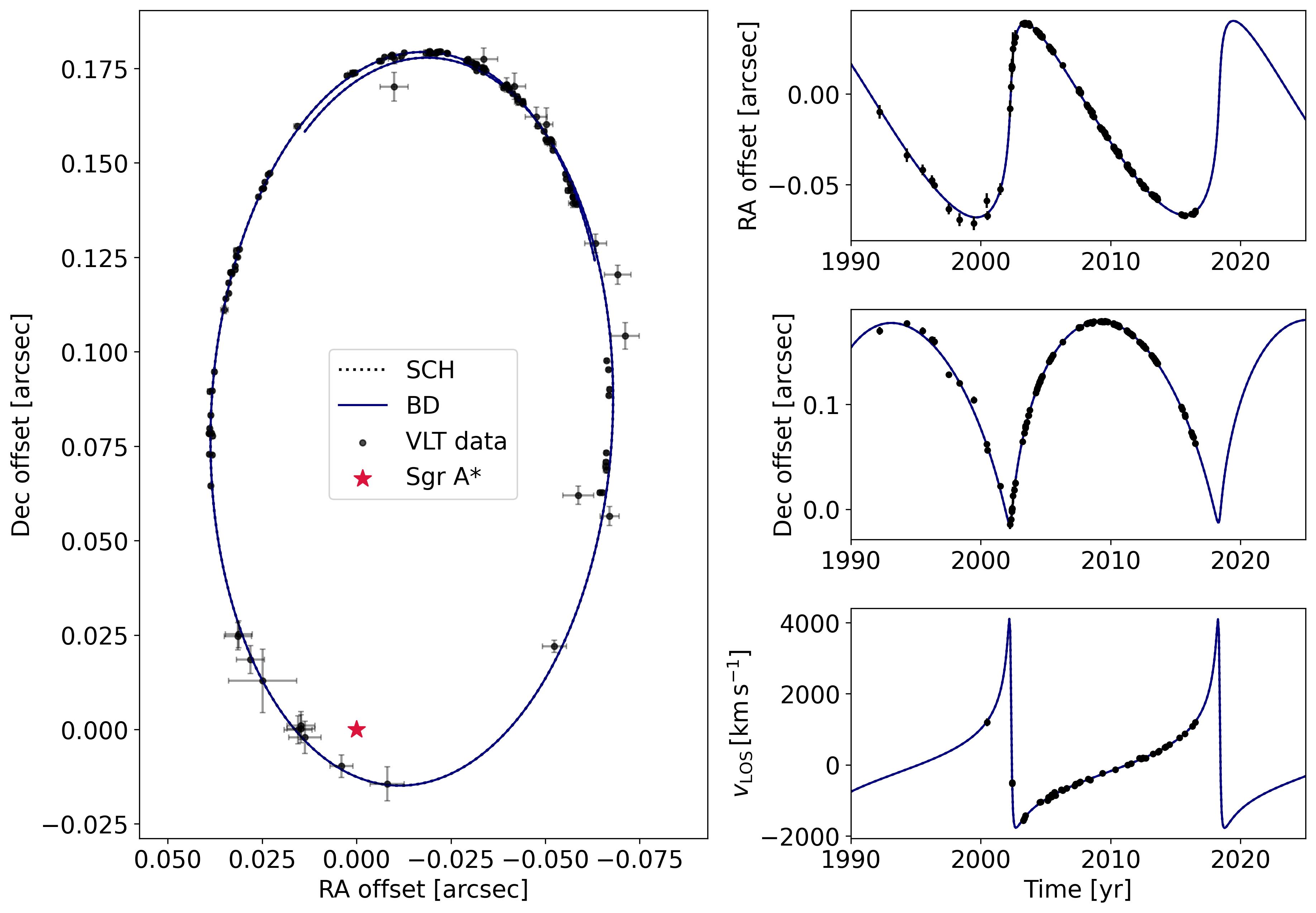}
    }\hfill
    \subfloat[RN and SCH spacetime\label{RNfit}]{\includegraphics[width=0.48\textwidth]{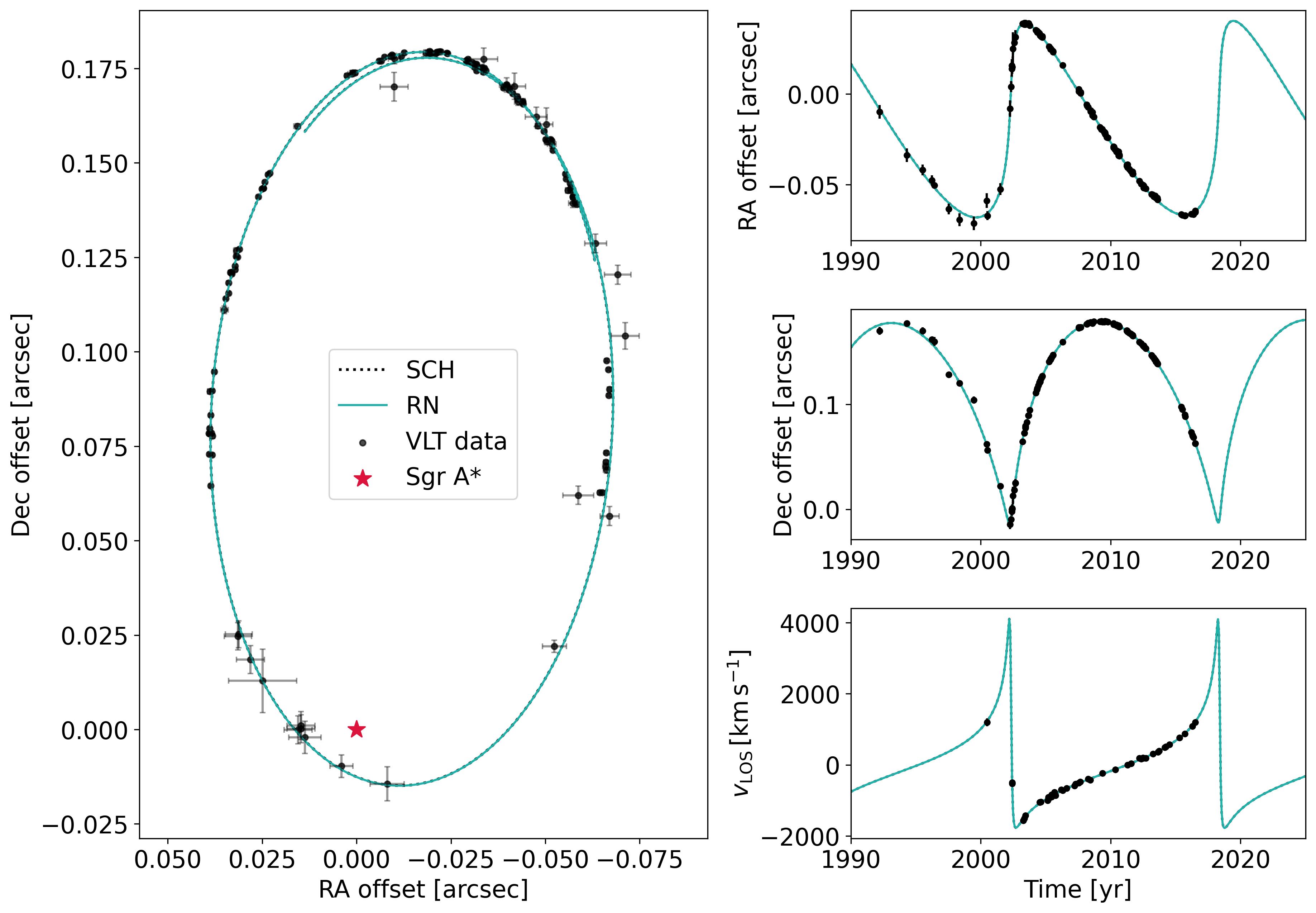}
}
\caption{This figure shows the combined astrometric and spectroscopic orbital fits of the S2 star using 24 years of VLT observations. The data are fitted within the Reissner--Nordström (RN) and Bardeen (BD) spacetimes and compared with the Schwarzschild (SCH) spacetime, which is shown by the dotted curve. For each spacetime, the left panel displays the best-fit sky-projected orbit, while the right panel shows the corresponding astrometric offsets and line-of-sight radial velocity measurements together with the model predictions.
}

\label{fig:orbitFit}
\end{figure*}
In this section, we present the results of our relativistic orbital analysis of the S2 star in different compact object spacetimes using the VLT astrometric and spectroscopic data. The fitting procedure includes sky-plane positional offsets and LoS velocity measurements, and all orbital integrations are performed fully relativistically, incorporating R\o mer time-delay and relativistic redshift effects.

The best-fitting orbital solutions in the Schwarzschild, RN and BD spacetime models are shown in Figure~\ref{fig:orbitFit}. For these representative cases, we plot the fully relativistic sky-projected orbits and the corresponding spectroscopic redshift curves. In all three models, the predicted trajectories and redshift signals are consistent with the observational data within the $2\sigma$ uncertainties. Although we only display results for Schwarzschild, RN and BD black holes for clarity, the remaining spacetime models considered in this work yield similar orbital fits to the S2 data.

\begin{figure*}
\centering
{\includegraphics[width=\linewidth]{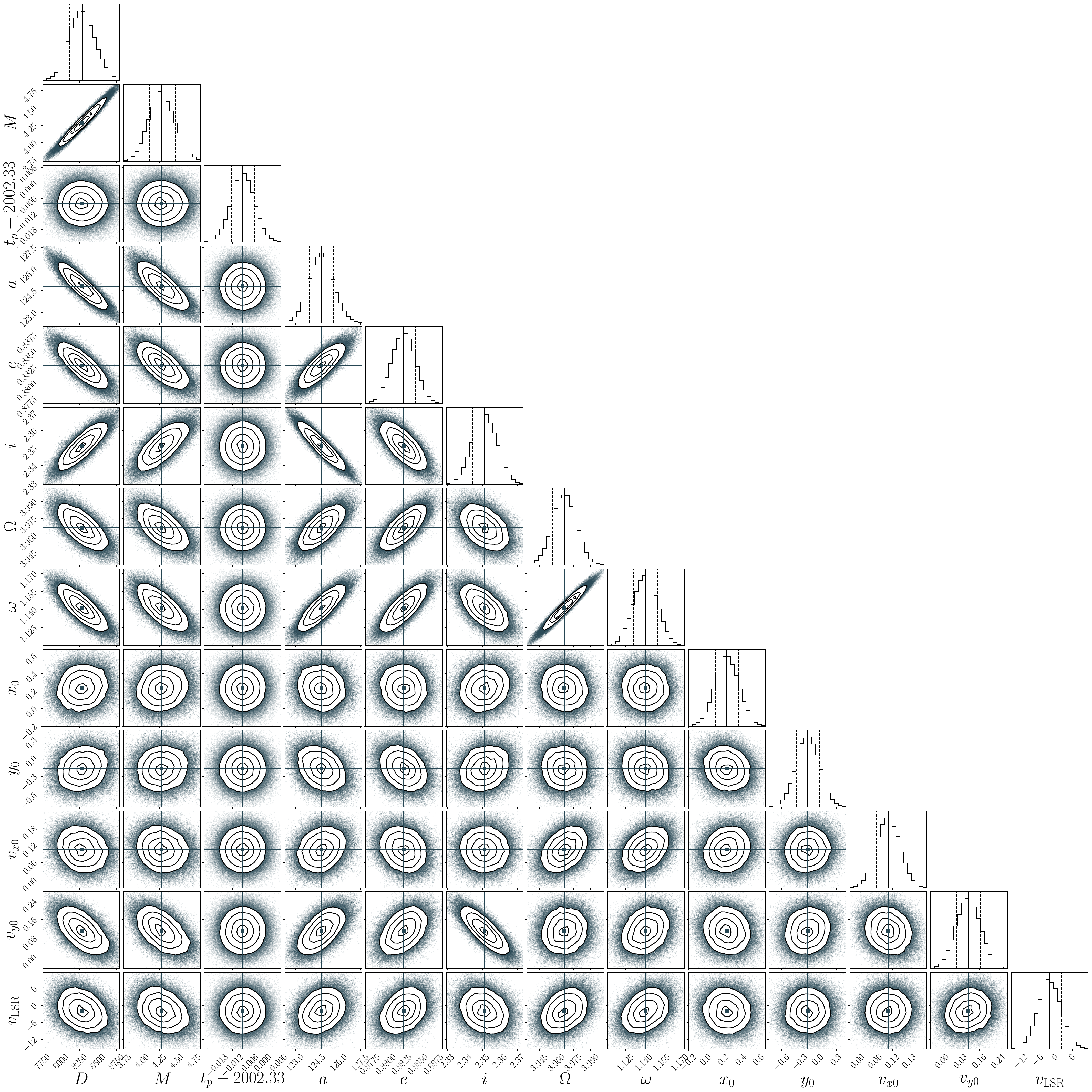}}
\caption{This figure shows the corner plot of the Schwarzschild model obtained from an MCMC analysis of the combined astrometric and spectroscopic S2 data. The contours indicate the $68\%$ and $95\%$ credible intervals. Best-fit values correspond to the posterior medians. The reduced chi-square is $\chi^2_\nu = 0.943$.}
\label{cornerSCH}
\end{figure*}

\begin{figure*}
\centering
{\includegraphics[width=\linewidth]{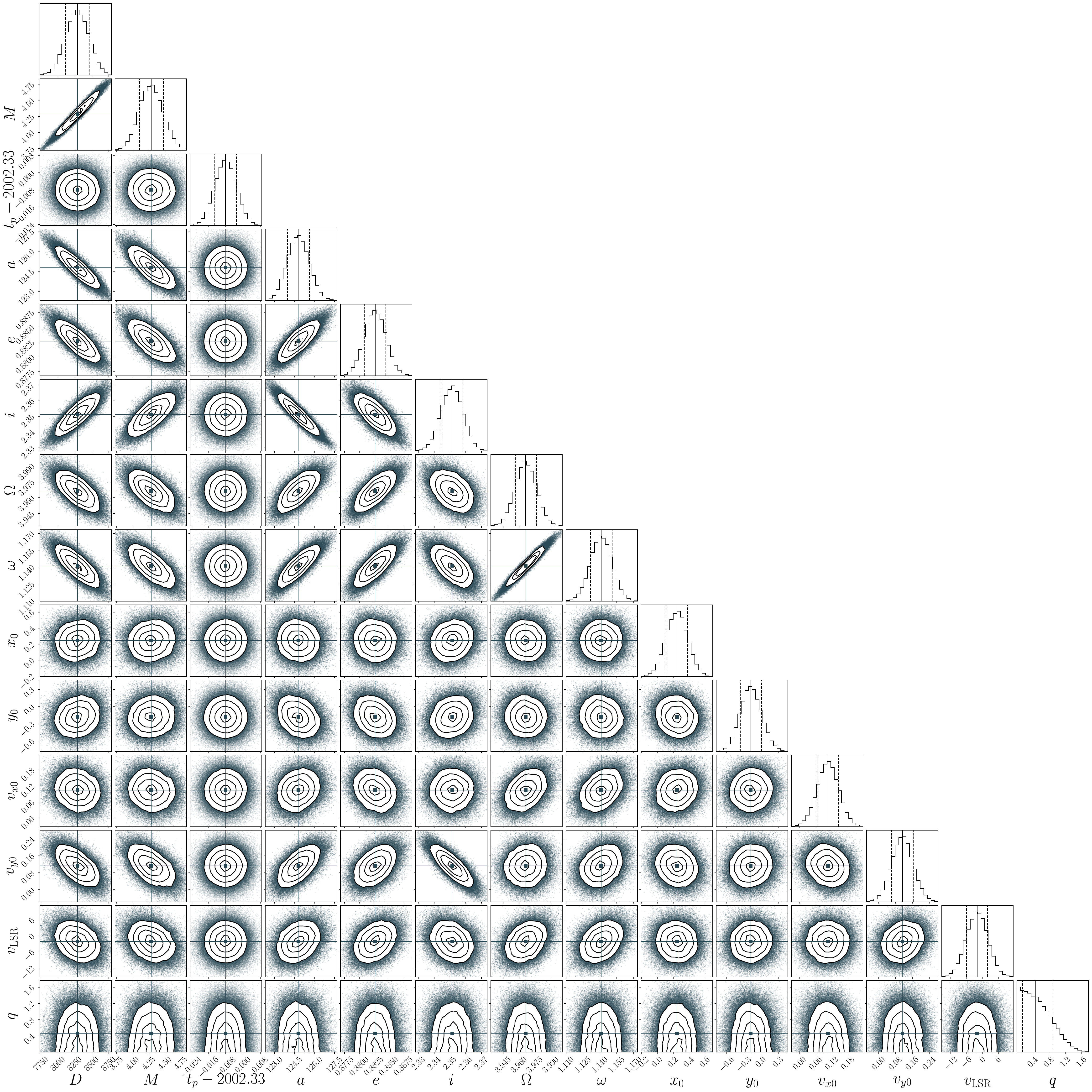}}
\caption{This figure shows the corner plot of the RN model obtained from an MCMC analysis of the combined astrometric and spectroscopic S2 data. The contours indicate the $68\%$ and $95\%$ credible intervals. Best-fit values correspond to the posterior medians. The reduced chi-square is $\chi^2_\nu = 0.946$.}
\label{cornerRN}
\end{figure*}

\begin{figure*}
\centering
{\includegraphics[width=\linewidth]{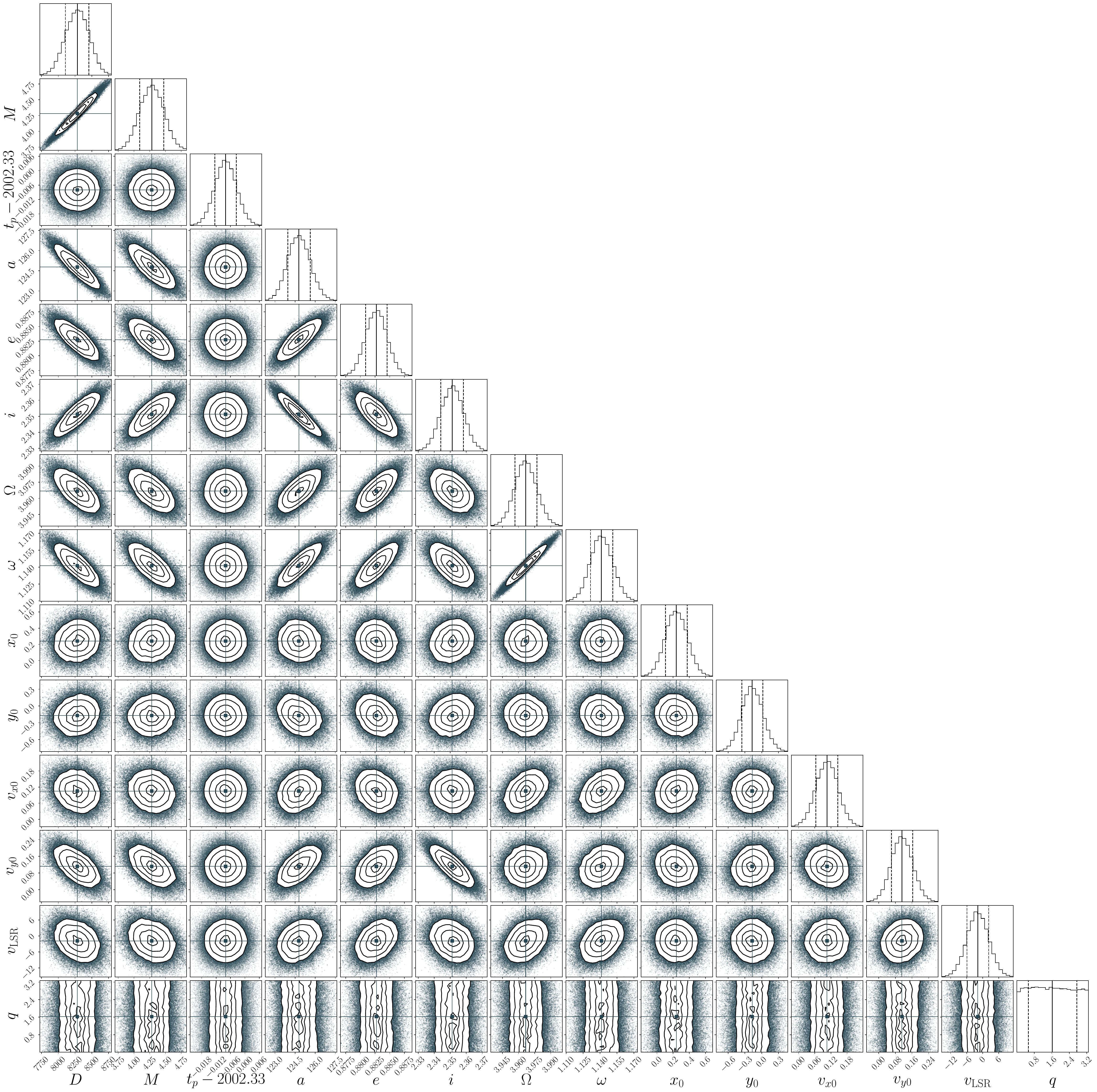}}
\caption{This figure shows the corner plot of the BD model obtained from an MCMC analysis of the combined astrometric and spectroscopic S2 data. The contours indicate the $68\%$ and $95\%$ credible intervals. Best-fit values correspond to the posterior medians. The reduced chi-square is $\chi^2_\nu = 0.946$.}
\label{cornerBD}
\end{figure*}

The marginalized posterior distributions and parameter correlations obtained from the Bayesian analysis of the S2 orbit in the Schwarzschild, RN and BD black hole spacetimes are shown in Figures~\ref{cornerSCH}, \ref{cornerRN}, and \ref{cornerBD}. These corner plots display the posterior distributions for the full set of orbital and spacetime parameters inferred from the current dataset, with the inner-most, middle and outermost regions indicating the $1\sigma$ (68\%), $2\sigma$ (95\%) and $3\sigma$ (99.7\%) credible intervals, respectively. The corresponding best-fit values and $2\sigma$ credible intervals are given in Table~\ref{tab:orbital_parameters}, and the associated $\chi^2$ statistics are summarized in Table~\ref{tab:compare}, while the adopted prior ranges for all fitted parameters are listed in Table~\ref{tab:prior}. We present corner plots only for the Schwarzschild, RN and BD black hole models, since the remaining spacetimes exhibit posterior structures that are qualitatively similar to these representative cases. For all parameters except the charge-like parameter $q$, the posteriors are approximately Gaussian and nearly identical across models. The posterior of $q$ exhibits noticeable non-Gaussian features in several cases, reflecting parameter degeneracies and weaker constraints from the current S2 dataset. Correspondingly, the marginalized likelihood for $q/M$ is not uniform across models and is often highly uneven, as shown in Figure~\ref{fig:mcmc_constraint}, due to the non-Gaussian nature of the $q$ posteriors. We further performed additional analyses with extended prior ranges for $q$ up to $q \leq 200\,M$ and find that, under such broad priors, the $q$ posterior approaches a Gaussian shape for all models; however, the resulting $1\sigma$ and $2\sigma$ bounds correspond to values that are very large and observationally excluded by existing black hole shadow constraints \cite{Vagnozzi:2022moj}. The final $1\sigma$ and $2\sigma$ constraints on the generalized charge-like parameter $q/M$ adopted in this work are reported in Table~\ref{tab:q_constraints}.

We find that the RN and BD black hole geometries provide fits to the S2 data that are statistically indistinguishable from, or only weakly disfavored relative to, the Schwarzschild case within current observational uncertainties, according to the AIC and BIC values. This highlights the present degeneracy between standard, charged, and regular black hole models when constrained by stellar dynamics alone.

Beyond the Schwarzschild and RN spacetimes, we also present the results for the BD, HY, and SV regular black hole geometries, together with the JNW naked singularity model, in Table~\ref{tab:orbital_parameters}. We emphasize that the inferred parameter bounds must be interpreted according to the physical sector of each spacetime model, namely whether the corresponding parameter range describes a black hole geometry or a horizonless singularity configuration.
Among all the considered models, the RN singular black hole and the BD regular black hole provide the statistically best fits to the S2 orbital data and remain essentially indistinguishable from the Schwarzschild spacetime within the current observational precision. In particular, the bounds obtained for the BD spacetime lie entirely within its regular black hole sector. On the other hand, the HY and SV regular black hole models, as well as the JNW naked singularity spacetime, also yield acceptable fits to the observed S2 dynamics, but with comparatively larger $\Delta\mathrm{AIC}$ values, indicating statistically less favored fits relative to Schwarzschild, RN, and BD geometries.
Overall, the information criteria consistently favor the Schwarzschild, RN, and BD spacetimes within their respective allowed parameter domains. These conclusions are independently supported by both the Bayesian MCMC analysis and complementary analytical consistency conditions derived from the observed pericenter precession of the S2 star. Taken together, our results indicate that present day stellar dynamics observations are still insufficient to unambiguously distinguish between Schwarzschild black holes and several regular or horizonless compact object geometries within their observationally allowed bounds.

\section{Discussion and Conclusion}
\label{sec:conc}

In this work, we examined the extent to which stellar dynamics can discriminate between a Schwarzschild black hole and alternative compact object geometries at the GC. We performed fully relativistic orbit integrations and modeled all relevant observational effects. The resulting predictions were confronted with the most complete astrometric and spectroscopic dataset currently available for the S2 star. This approach complements EHT constraints by probing the spacetime through timelike geodesics over a wide radial range extending from hundreds to thousands of gravitational radii, rather than photon trajectories alone. A key result is that several spacetimes that are degenerate at the level of the black hole shadow also remain degenerate statistically and analytically when tested against present S2 data. In particular, the Schwarzschild, RN and BD geometries yield comparable likelihoods and information criteria. This shows that, at current observational precision, stellar orbit measurements alone cannot break certain spacetime degeneracies.

We note that, while this work was in preparation, a related study appeared on the arXiv presenting a similar analysis for several compact object spacetimes~\cite{2026MNRAS.tmp...47N},
including the Schwarzschild, Bardeen, Yukawa-like, JNW,
Horndeski, RN, and Einstein-Maxwell-dilaton geometries. In~\cite{2026MNRAS.tmp...47N}, the authors report that the Bardeen spacetime is statistically indistinguishable from the Schwarzschild solution when tested against the current S2 star data. While within our framework, we find that the RN and Bardeen spacetimes remain statistically indistinguishable from Schwarzschild, whereas
the HY, SV and JNW spacetimes remain slightly disfavored. This partially agrees with the results reported in Ref.~\cite{2026MNRAS.tmp...47N}. Some quantitative differences nevertheless arise between the two analyses, particularly in the treatment of the generalized charge-like parameters and the corresponding parameter ranges explored. Our analysis is performed using the publicly available \texttt{PyGRO} code~\cite{DellaMonica:2025npq}, which provides a fully relativistic treatment of timelike geodesics together with their astrometric and spectroscopic observables.
For the Bardeen spacetime, Ref.~\cite{2026MNRAS.tmp...47N} reported the constraint
$q/M \leq 17.13$
at the $1\sigma$ level. However, the extremal limit for the existence of a Bardeen black hole is
$q/M \leq \sqrt{16/27} \approx 0.77$~\cite{Vagnozzi:2022moj}, above which the spacetime corresponds to a globally regular horizonless compact object. In our analysis, we therefore focus on the physically admissible black hole regime and obtain
$q/M \leq 0.755$
at the $2\sigma$ level (see Table~\ref{tab:q_constraints}).In~\cite{2026MNRAS.tmp...47N}, the constraint on the Bardeen charge parameter is reported as
$q/M \leq 17.13$
at the $1\sigma$ level, which substantially exceeds the theoretically allowed range
$q/M \leq \sqrt{16/27} \approx 0.77$
required for the existence of a Bardeen black hole~\cite{Vagnozzi:2022moj}. For $q/M > \sqrt{16/27}$,
the spacetime no longer possesses horizons and instead describes a globally regular horizonless compact object. In our analysis, we adopt conservative priors on the charge parameter, allowing
$q \lesssim M$
(in units of $10^{6}M_{\odot}$), and nevertheless obtain a significantly tighter constraint,
$q/M \leq 0.755$
at the $2\sigma$ level (see Table~\ref{tab:q_constraints}), which lies within the theoretically allowed black hole regime.

Similarly, for the JNW spacetime, Ref.~\cite{2026MNRAS.tmp...47N} reported a $1\sigma$ bound of
$q/M \leq 33.71$,
whereas our analysis yields
$q/M \leq 0.983$
at the $2\sigma$ level, with the analytical analysis giving
$q/M \leq 0.965$.
For the RN spacetime, the constraint reported in Ref.~\cite{2026MNRAS.tmp...47N} is broadly consistent with previous analyses~\cite{Mishra:2023uxl}. In our analysis, both the analytical and MCMC constraints, together with EHT shadow bounds~\cite{Vagnozzi:2022moj}, favor the black hole regime
$q/M \leq 1$. Overall, the two studies lead to qualitatively similar conclusions for several common spacetimes, while differing quantitatively in the inferred bounds and parameter ranges considered.

Our results are also consistent with the conclusions of the parametrized post-Newtonian (PPN) analysis of the S2 orbit presented in~\cite{Saida:2024cjc}. In that work, the authors tested deviations from the Schwarzschild spacetime by fitting a spherically symmetric PPN metric to astrometric and spectroscopic observations of S2 using a $\chi^2$ framework. They found that while the best-fit PPN model yields a slightly lower reduced $\chi^2$ than the Schwarzschild case, the difference is too small to establish statistical significance with current data. This led them to conclude that present S2 observations do not allow a decisive discrimination between Schwarzschild and non-Schwarzschild geometries, and that more robust Bayesian methods are required. This conclusion closely supports our results. 

Despite considering a broader class of black hole and horizonless spacetimes, and employing fully relativistic orbit integration together with Bayesian inference and information criteria, we find that several alternative geometries remain degenerate with the Schwarzschild spacetime at current observational precision. Also note that we did not perform the S2 star orbit analysis in the JMN-1 spacetime. However, since the photon sphere is located at $r \gtrsim 3M$, the JMN-1 spacetime is smoothly matched to the Schwarzschild geometry beyond this radius. Therefore, the orbital motion of the S2 star, which occurs well outside this region, would be indistinguishable from the Schwarzschild case~\cite{Saurabh:2023otl}. Our results remain consistent with the shadow image of Sgr A*, while respecting the external Schwarzschild matching.

In~\cite{Cadoni:2022vsn}, authors proposed a non-singular black hole spacetime with a de Sitter core and a deformation parameter $\ell$, showing that S2 orbital data place a strong upper bound $\ell \lesssim 0.47\,GM$ while not ruling out regular black holes. This also supports our finding that current stellar dynamics constrain, but do not uniquely exclude, regular geometries at the GC. These results indicate that any deviations from the Schwarzschild metric, if present, lie below the sensitivity threshold of existing S2 data. Both analyses therefore emphasize the need for improved astrometric accuracy and more sophisticated statistical frameworks to enable decisive tests of the spacetime geometry at the GC.

These results highlight the synergistic roles of stellar dynamics and horizon-scale imaging as probes of the GC spacetime. While the EHT shadow constrains the properties of the photon region, stellar orbits probe the gravitational field through the motion of massive particles across a broad radial range. The persistence of degeneracies across both observables emphasizes the need for improved observational precision and additional, independent probes to test deviations from the Kerr paradigm. At present, S2 data place meaningful constraints on several alternative compact object models, but do not uniquely determine the causal structure of Sgr A*. Future work could extend this analysis by including the spin of the central compact object. A fully relativistic treatment of frame-dragging effects on S2 and other S-star orbits would enable accurate projection onto the observer’s sky, allowing direct fitting to astrometric and spectroscopic data. Combined with future improvements in astrometric accuracy, longer observational baselines, and additional short period stars, such analyses could significantly strengthen the ability to distinguish black holes from their alternatives at the GC.

\acknowledgments{P. Bambhaniya, and Elisabete M. de Gouveia Dal Pino acknowledge support from the São Paulo Research Foundation (FAPESP) under grants No. 2024/09383-4. G. H. Vicentin and Elisabete M. de Gouveia Dal Pino acknowledge support from the São Paulo Research Foundation (FAPESP) under grants No. 2023/10590-1. Elisabete M. de Gouveia Dal Pino also acknowledges support from FAPESP under grant No. 2021/02120-0 and from the Brazilian National Council for Scientific and Technological Development (CNPq) under grant No. 308643/2017-8. R. Della Monica acknowledges financial support provided by FCT – Fundação para a Ciência e a Tecnologia, I.P., through the ERC-Portugal program Project ``GravNewFields''. R. Della Monica also thanks the Fundação para a Ciência e Tecnologia (FCT), Portugal, for the financial support to the Center for Astrophysics and Gravitation (CENTRA/IST/ULisboa) through grant No.~\href{https://doi.org/10.54499/UID/PRR/00099/2025}{UID/PRR/00099/2025} and grant No.~\href{https://doi.org/10.54499/UID/00099/2025}{UID/00099/2025}. P. Bambhaniya thanks Saurabh and Meet Vyas for useful discussions.}

\begin{widetext}
\begin{table*}[h]
    \centering
    \small
    \renewcommand{\arraystretch}{1.6}
    \setlength{\tabcolsep}{4pt}
    \caption{Best-fit orbital and physical parameters for different compact object models.}
    \begin{tabular}{|c|c|c|c|c|c|c|}
        \toprule
        \hline
        \hline
        \textbf{Parameters} 
        & \textbf{SCH} 
        & \textbf{RN} 
        & \textbf{BD} 
        & \textbf{HY} 
        & \textbf{SV}
        & \textbf{JNW}  \\
        \midrule
        \hline

        $D$ (kpc) 
        & $8.28^{+0.36}_{-0.33}$ 
        & $8.29^{+0.33}_{-0.34}$ 
        & $8.28^{+0.36}_{-0.35}$ 
        & $8.28 \pm 0.34$ 
        & $8.29^{+0.33}_{-0.34}$ 
        & $8.28^{+0.34}_{-0.33}$ \\

        \hline
        $M$ ($10^6\,M_\odot$) 
        & $4.28^{+0.40}_{-0.35}$ 
        & $4.29^{+0.38}_{-0.36}$ 
        & $4.29^{+0.40}_{-0.38}$ 
        & $4.28^{+0.38}_{-0.36}$ 
        & $4.29^{+0.38}_{-0.36}$ 
        & $4.29^{+0.38}_{-0.35}$ \\

        \hline
        $t_p - 2002$ (yr) 
        & $0.322^{+0.010}_{-0.009}$ 
        & $0.322^{+0.010}_{-0.009}$ 
        & $0.322 \pm 0.010$ 
        & $0.322 \pm 0.009$ 
        & $0.322 \pm 0.009$ 
        & $0.323^{+0.009}_{-0.010}$ \\

        \hline
        $a$ (mas) 
        & $124.8^{+1.7}_{-1.6}$ 
        & $124.8^{+1.7}_{-1.6}$ 
        & $124.8^{+1.8}_{-1.6}$ 
        & $124.8 \pm 1.6$ 
        & $124.7 \pm 1.6$ 
        & $124.8^{+1.7}_{-1.6}$ \\

        \hline
        $e$ 
        & $0.883 \pm 0.004$ 
        & $0.883 \pm 0.004$ 
        & $0.883 \pm 0.004$ 
        & $0.883 \pm 0.004$ 
        & $0.883 \pm 0.004$ 
        & $0.883 \pm 0.004$ \\

        \hline
        $i$ (rad) 
        & $2.351 \pm 0.014$ 
        & $2.351 \pm 0.014$ 
        & $2.351 \pm 0.014$ 
        & $2.351^{+0.014}_{-0.013}$ 
        & $2.351 \pm 0.014$ 
        & $2.351 \pm 0.014$ \\

        \hline
        $\Omega$ (rad) 
        & $3.967 \pm 0.021$ 
        & $3.966 \pm 0.020$ 
        & $3.967 \pm 0.021$ 
        & $3.967 \pm 0.021$ 
        & $3.966^{+0.020}_{-0.021}$ 
        & $3.967^{+0.020}_{-0.021}$ \\

        \hline
        $\omega$ (rad) 
        & $1.141 \pm 0.020$ 
        & $1.141 \pm 0.019$ 
        & $1.141 \pm 0.020$ 
        & $1.141^{+0.019}_{-0.020}$ 
        & $1.141^{+0.019}_{-0.020}$ 
        & $1.141^{+0.019}_{-0.020}$ \\

        \hline
        $x_0$ (mas) 
        & $0.24 \pm 0.27$ 
        & $0.25 \pm 0.27$ 
        & $0.24^{+0.28}_{-0.27}$ 
        & $0.24^{+0.28}_{-0.27}$ 
        & $0.23 \pm 0.27$ 
        & $0.24 \pm 0.27$ \\

        \hline
        $y_0$ (mas) 
        & $-0.17 \pm 0.38$ 
        & $-0.18^{+0.39}_{-0.37}$ 
        & $-0.17 \pm 0.38$ 
        & $-0.17 \pm 0.38$ 
        & $-0.16^{+0.38}_{-0.39}$ 
        & $-0.17 \pm 0.38$ \\

        \hline
        $v_{x0}$ (mas\,yr$^{-1}$) 
        & $0.105^{+0.082}_{-0.080}$ 
        & $0.105^{+0.082}_{-0.083}$ 
        & $0.106^{+0.082}_{-0.081}$ 
        & $0.106^{+0.080}_{-0.082}$ 
        & $0.105^{+0.083}_{-0.080}$ 
        & $0.105 \pm 0.082$ \\

        \hline
        $v_{y0}$ (mas\,yr$^{-1}$) 
        & $0.11 \pm 0.10$ 
        & $0.11 \pm 0.10$ 
        & $0.11 \pm 0.10$ 
        & $0.114^{+0.099}_{-0.10}$ 
        & $0.11 \pm 0.10$ 
        & $0.11 \pm 0.10$ \\

        \hline
        $v_{\rm LSR}$ (km\,s$^{-1}$) 
        & $-2.0^{+8.3}_{-8.0}$ 
        & $-2.1 \pm 8.1$ 
        & $-2.0^{+8.1}_{-8.0}$ 
        & $-1.9^{+8.4}_{-8.3}$ 
        & $-2.0^{+8.1}_{-8.3}$ 
        & $-2.0 \pm 8.2$ \\

        \hline
        $q$ ($10^6\,M_\odot$) 
        & -- 
        & $0.47^{+0.85}_{-0.45}$ 
        & $1.6^{+1.6}_{-1.5}$ 
        & $1.7^{+1.5}_{-1.6}$ 
        & $4.6^{+4.5}_{-4.4}$ 
        & $2.2^{+1.9}_{-2.1}$ \\

        \bottomrule
        \hline
    \end{tabular}
    \label{tab:orbital_parameters}
\end{table*}

\begin{table}[h]
    \centering
    \small
    \renewcommand{\arraystretch}{1.4}
    \setlength{\tabcolsep}{10pt}
    \caption{$\chi^2_{\rm min}$ values and statistical comparison of all models with respect to Schwarzschild model}
    \begin{tabular}{|c|c|c|c|c|c|c|c|}
        \toprule
        \hline
        \hline
        \textbf{Models} 
        & \textbf{$\chi^2_{\rm min}$} 
        & \textbf{$\chi^2_{v}$} 
        & \textbf{AIC} 
        & \boldsymbol{$\Delta$}\textbf{AIC} 
        & \textbf{BIC} 
        & \boldsymbol{$\Delta$}\textbf{BIC} 
        & \textbf{Evidence} \\
        \midrule
        \hline 
        SCH & 303.700 & 0.943 & 329.700 & -- & 379.283 & -- & --\\
        \hline
        RN & 303.589 & 0.946 & 331.589 & 1.889 & 384.987 & 5.703 & Indistinguishable to Weak\\
        \hline
        BD & 303.644 & 0.946 & 331.644 & 1.944 & 385.042 & 5.758 & Indistinguishable to Weak\\
        \hline
        HY & 303.860 & 0.947 & 331.860 & 2.160 & 385.258 & 5.975 & Weak\\
        \hline
        SV & 303.749 & 0.946 & 331.749 & 2.049 & 385.147 & 5.863 & Weak\\
        \hline
        JNW & 303.764 & 0.946 & 331.764 & 2.064 & 385.162 & 5.878 & Weak\\
        
        \bottomrule
        \hline
    \end{tabular}
    \label{tab:compare}
\end{table}
\end{widetext}

\nocite{*}
\bibliographystyle{apsrev4-2}
\bibliography{references}

\clearpage

\end{document}